\documentclass[twocolumn,10pt]{article}
\usepackage{times} 
\usepackage{balance} 
\usepackage{bbm}
\usepackage[utf8]{inputenc}
\usepackage[margin=1in]{geometry}
\usepackage[labelfont=bf, skip=5pt, font=small]{caption}
\usepackage[svgnames]{xcolor}
\usepackage{soul}
\usepackage{graphicx}
\usepackage{fancyhdr}
\usepackage{amsmath}
\usepackage{hyperref} 
\usepackage{amssymb}
\usepackage[style=nature, articletitle=true]{biblatex}
\usepackage{indentfirst}
\usepackage{subcaption}
\usepackage{datetime}
\usepackage{titlesec} 
\usepackage{epstopdf}
\usepackage{pdfpages}
\usepackage{pdflscape}
\usepackage{microtype}
\usepackage{cancel}
\usepackage[switch]{lineno}
\usepackage{cuted}

\usepackage{authblk}
\usepackage{orcidlink}

\let\oldsim\sim
\renewcommand{\sim}{{\oldsim}}
\DeclareCaptionFont{myblue}{\color{DarkBlue}}
\title{\bfseries Passage of particles through matter and the effective straggling-function: High-fidelity accelerated simulation via Physics-Informed Machine Learning}

\author[1]{Oleksandr {Borysov}}
\author[1]{Rotem {Dover}\thanks{Corresponding author: rotem.dover@weizmann.ac.il}\orcidlink{0009-0001-1491-3834}}
\author[1]{Eilam {Gross}}
\author[1]{Nilotpal {Kakati}}
\author[1]{Noam {Tal Hod}}

\affil[1]{Department of Particle Physics and Astrophysics, Weizmann Institute of Science, 234 Herzl Street, Rehovot, 7610001, Israel}

\date{} 

\begin{document}

\twocolumn[
\begin{@twocolumnfalse}
\maketitle 
\begin{abstract}
High-fidelity simulation of particle–matter interactions provides the essential theoretical reference for diverse physics disciplines, yet generating synthetic datasets at the scale of current and future experiments has become prohibitive.
Here, we introduce PHIN-GAN, a novel physics-informed generative adversarial network designed to address this challenge.
We derive a set of analytical probability density functions, that effectively describe the ``straggling function'' identified with Landau.
For the first time, this enables their continuous evaluation across the entire phase-space.
These analytical forms are leveraged to enforce a parametric distribution-level learning objective. 
Rooted in first principles, PHIN-GAN offers a generalizable, interpretable and scalable proof-of-concept approach for a lossless generative model that maintains the high fidelity of the standard-bearer for simulating such interactions, namely GEANT4, at a fraction of the computational cost.
\end{abstract}
\vspace{2em} 
\end{@twocolumnfalse}
]

\section{Introduction}\label{sec:introduction}
\setlength{\parskip}{1em}
\setlength{\parindent}{0.5em}

High-fidelity simulations of elementary particle interactions with matter are a cornerstone of modern research across diverse scientific disciplines, including High-Energy Physics (HEP), medical physics, plasma and laser physics, accelerator physics, nuclear and fusion physics, archaeology, industrial applications, and space science.
Within these fields, simulation plays a consistent and essential role: providing the theoretical reference required to model and interpret experimental data. 
This is typically achieved using classical Monte-Carlo (MC) methods, which enable modeling of the underlying physics through the generation of large synthetic datasets of discrete random events.

Over four decades now, the standard-bearer for simulating MC events of particle-matter interactions is GEANT4~\cite{agostinelli2003geant4, 1610988, ALLISON2016186}. 
This state-of-the-art toolkit encapsulates extensive theoretical and experimental research carried out by thousands of scientists.
It enables thorough understanding of physics processes through the confrontation of simulation with data.
Indeed, its physics modeling matches an astonishingly wide range of experimental results with an unprecedented precision.
GEANT4's far-reaching impact is evident in its widespread adoption, reflected by over 60,000 citations of its official core publications.
However, the aggregate cost of GEANT4's meticulous modeling is becoming prohibitive given the massive datasets required by current and future physics programs worldwide.
Addressing this necessitates a fundamental re-evaluation of particle transport through matter computation.
Consequently, the GEANT4 road-map identifies a critical need for a 10- to 100-fold acceleration to reconcile the physics goals of future experiments with sustainable computing budgets~\cite{g4fastsim}.

The growing availability of Graphics Processing Units (GPUs) offers a pathway to accelerate the respective simulations through massive parallelism.
To exploit this, the community is primarily pursuing two distinct strategies:
developing fast generative surrogates that bypass GEANT4 and adapting parts of GEANT4's physics engine to run on GPUs.
The disparity between the rapid approximations of generative models and the bounded speedups of full-fidelity approaches highlights the need for a hybrid framework—one that leverages the fast generative models not to replace the physics, but rather to accelerate the individual steps of the transport engine itself.

We propose a novel physics-informed generative adversarial network, denoted hereafter as `PHIN-GAN', for simulating fundamental particle-matter interactions. 
This proof-of-concept work introduces a twofold novelty: first, we derive a set of analytical probability density functions (PDFs) that effectively describe the ``straggling function'' commonly identified with Landau~\cite{Landau:216256}.
For the first time, this effective straggling function can be evaluated continuously across the entire phase-space. 
Second, we leverage these analytical forms to enforce a parametric \textit{distribution-level learning objective}.
By rooting the methodology in first principles, we establish an inherently generalizable, scalable, and interpretable proof-of-concept (POC) approach for a lossless generative model that preserves the high fidelity of GEANT4.
Extending this framework to include additional processes, materials, and particle species--ultimately aiming to preserve the comprehensive fidelity of GEANT4--remains an important direction for future work.

This paper is structured as follows: Sec.~\ref{sec:state-of-the-art} characterizes the computational bottleneck in greater detail using a concrete case study, and reviews the emerging strategies to address it.
We then continue to describe briefly the GEANT4 machinery focusing on the most dominant subset of electromagnetic (EM) interactions as well as the mathematical framework used to derive the related physics PDFs.
This foundation leads to the introduction of PHIN-GAN, detailing its architecture, training dataset, and physics-informed learning objective. 
Finally, we validate the model through rigorous performance benchmarks and discuss the path toward extending this framework to full particle cascade simulations.

\section{State of the art}\label{sec:state-of-the-art}
\setlength{\parskip}{1em}
\setlength{\parindent}{0.5em}
Before introducing our new approach, it is essential to first discuss the problem to tackle, as well as other emerging strategies that attempt to cope with the expected escalating dataset sizes of current and future experimental physics programs.

\noindent\textbf{The computational problem:} While indispensable across the broad landscape of scientific domains mentioned earlier~\cite{perl2012topas, jan2004gate, incerti2010geant4, morishima2017discovery, hohlmann2009geant4, atwood2009large}, GEANT4's rigorous modeling of particle trajectories incurs a substantial computational cost.
This high resource consumption is particularly acute in HEP~\cite{aad2008atlas, cms2008cms, aubert2002babar, belle2002belle, schael2006precision}, where the imperative to distinguish rare signals from typically large backgrounds demands an ever-increasing statistical precision.
In the ATLAS experiment at the Large Hadron Collider (LHC), for example, simulating particle interactions consumed $\sim$30\% of ATLAS's gigantic computing resources during Run~2 (2015-2018)~\cite{CERN-LHCC-2020-015, CERN-LHCC-2022-005}, corresponding to the generation of approximately 52 billion proton-proton collisions.
For reference, the worldwide LHC computing grid (WLCG) combines, as of 2025, about 1.4 million computer cores from over 170 sites in 42 countries~\cite{Campana:2018, CERN_WLCG}.
This LHC example is not unique; current experiments face comparable challenges, where future experiments are expected to face even greater ones~\cite{apollinari2016high, abada2019fcc, khalek2022science, black2024muon}.

\noindent\textbf{Emerging strategies:} The community is primarily pursuing two distinct strategies:
utilizing fast generative surrogates to bypass GEANT4 or adapting parts of GEANT4's physics engine to run on GPUs.
Each of these strategies has its pros and cons.
The first is typically implemented with Deep Generative Models (DGMs) (see~\cite{Hashemi_2024} for a comprehensive review).
While achieving orders-of-magnitude speedups, DGMs typically bypass the micro-physics description to emulate final high-level responses.
Consequently, they are often task-specific, sacrificing granular detail for macroscopic approximations without directly incorporating the underlying physics.
In the second strategy, projects like \textit{Celeritas}~\cite{johnson2024celeritas} maintain full fidelity by implementing parts of GEANT4 on GPUs.
However, algorithmic overheads limit the observed speedups to factors of $1.8$--$2.3$~\cite{johnson2024celeritas}, which is insufficient to meet the demands depicted above.

\section{GEANT4 EM Interactions in a nutshell}\label{sec:geant4-simulation-framework}
\begin{figure}[!ht]
\centering
\includegraphics[width=0.48\textwidth]{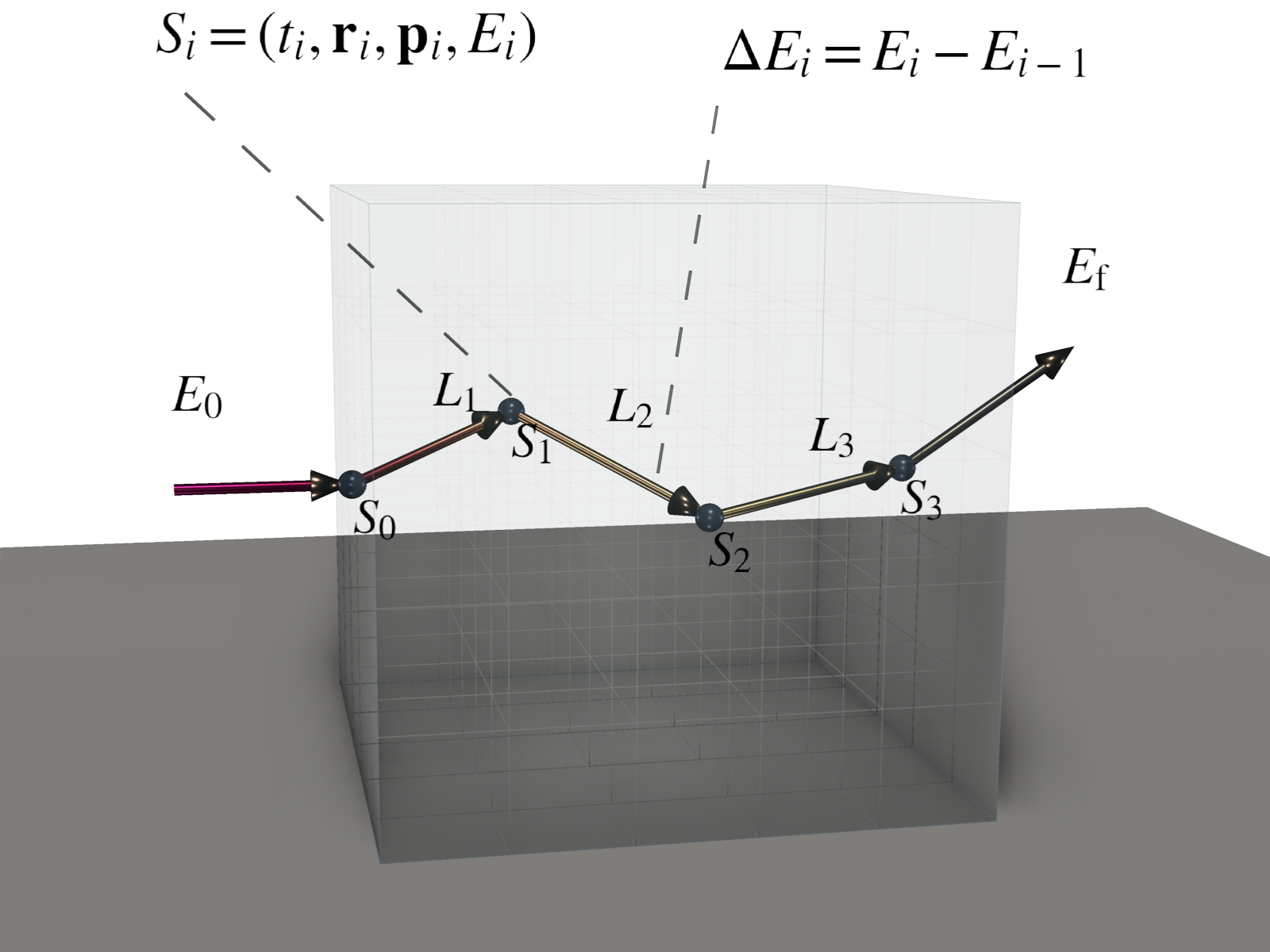}
\caption{A conceptual sketch of a particle’s propagation through a sequence of steps in GEANT4.
At each step, the particle’s state (denoted as $S$)—including its position ($\textbf{r}$), momentum ($\textbf{p}$), and energy ($E$) at a given timestamp ($t$)—is updated.
This state evolves due to various physical processes, which include energy-loss, scattering, and the generation of
secondary particles.}
\label{fig:steps_sketch}
\hfill
\end{figure}

\setlength{\parskip}{1em}
\setlength{\parindent}{0.5em}
At the core of GEANT4’s framework, particle propagation is achieved by sequentially chaining discrete stochastic \textit{steps} to form a coherent \textit{trajectory}.
This stepping process serves as the fundamental building block.
The conceptual propagation of a particle and its momentary states along a single step is presented in Fig.~\ref{fig:steps_sketch}.
At the beginning of each step, a pre-step point is defined with the current position, momentum, energy and elapsed
time from beginning of the simulation.
This point represents the particle's state before the step.
GEANT4 evaluates the cross-sections (probabilities) of all applicable processes given the particle's state and the environment, and subsequently draws a random step length for each process.
The physics step process is selected and executed as the process with the shortest step length.
This alters the particle's state and may also yield secondary particles, which would be tracked separately.
Energy-loss can occur continuously along the step, discretely at its ending, or a combination of both.

In this POC work, we focus only on excitation and ionization collisions of charged particles with atomic electrons, as they constitute a significant fraction of the overall interactions in most typical simulation chains.
As long as the energy transfer at collisions is lower than a threshold defined (indirectly) by the user's settings, the primary particle propagates continuously.
During this stage, GEANT4 deposits the energy from these sub-threshold interactions locally and does not generate separate trajectories for these electrons.
A collision is considered as a discrete process, when the energy transfer to the atomic electron is above the threshold.
As a result of such process, this electron is tracked as a secondary to the primary particle.

We simulate the interactions of 100~MeV primary protons propagating through a 5~cm aluminum bulk depositing all of their energy.
These choices allow us to avoid the high-energy radiative processes and focus on the excitation and ionization alone.
The registered EM process class for ionization processes is \texttt{G4hIonisation}.
In this work, we turn off multiple-scattering (MSC), to focus on the energy-loss.
Doing so ties the scattering angle entirely to the collision that yields the secondary electron.
This leads to a highly non-trivial angular distribution that is significantly more difficult for a neural network to learn compared to the case where the MSC process is turned on and the distribution becomes dominated by it.
We leave the inclusion of MSC process for a future work, but note that preliminary tests indicate that it can be easily modeled as long as the two angles of the secondary and the primary particles are treated separately.
The default model for energy-loss fluctuations is \texttt{G4UniversalFluctuation}.
This is an advanced and precise approach developed for GEANT4 rooted in the seminal work of Lassila-Perini and Urb\'an~\cite{lassila1995energy} known as the ``Urb\'an model''.
It is based on an MC framework that simulates energy-loss without an underlying PDF, capturing the complex stochastic processes involved.

Originally, the fluctuations in the energy-loss were understood through Landau’s theory, formulated in 1944~\cite{Landau:216256}.
Landau’s work introduced the now-famous `straggling function' to describe the asymmetric energy-loss profile resulting
from many low-energy collisions with occasional high-energy transfers.
Following Landau’s initial work, there were many attempts to improve and generalize the description, see for
example~\textcite{osti_4311507} and~\textcite{bichsel1988straggling}.
These theories are valid for long steps, where many collisions occur, allowing energy-loss to be treated as an average process.
At short steps, fewer collisions and the significance of atomic structure require a more detailed treatment of the fluctuations.
While these theories are very successful, they do not provide a unified practical framework allowing to easily write down the PDFs for any kind of incident particle, media or any point in the phase-space (particle energy and step length).
The calculations are very intense, should be repeated for different configurations and they are only applicable in a small
region of the phase-space.
Moreover, it is possible to identify that the development of these efforts was gradually “stopped” around the time when
the first versions of GEANT were introduced during the mid 1970s and early 1980s.
This trend became even stronger with the introduction of the Urb\'an effective model in the mid 1990s~\cite{lassila1995energy}.
Presently, the models implementated in GEANT4 are considered to be the most accurate description available for the
interaction of radiation and matter, albeit in the form of MC rather than analytical PDFs\@.
The production of large scale simulation is deemed not practical if one is to base the generation directly on the
early theories, while the effective models provide the needed accuracy.
To address this gap, GEANT4 implements a more detailed fluctuation model, improving upon the Landau straggling function by
simulating energy-losses directly via MC methods.
This modern approach surpasses the traditional analytical methods by offering a highly accurate representation of
the actual, step-wise energy-loss, especially in thin-layer media (and in the GEANT4 case, short steps).
As a result, GEANT4 provides a more precise and reliable model for energy-loss in a wide range of materials,
enabling experiments to simulate and interpret particle interactions with unprecedented fidelity.
The agreement between GEANT4 simulations and a vast range of experimental data is remarkably robust, consistently
validating its accuracy across diverse scenarios.

To incorporate the effective straggling function of charged particles' energy-loss fluctuations into the learning process of a
generative model, a complete expression for $\mathbb{P}(\Delta E|E)$ is required, where $\mathbb{P}$ denotes the PDF and
$E$ and $\Delta E$ are the energy and energy-loss of the particle respectively.
Besides the explicit energy dependence, this PDF depends also on the particle species and the medium traversed.
In the following section, we present the first derivation of an analytical model of the effective straggling function for charged particle energy-loss fluctuations, based on
\texttt{G4UniversalFluctuation} implementation, which is in turn based on the modified Urb\'an model.
To date, we have successfully derived $\mathbb{P}(\Delta E|E, L)$, where $L$ is the step length.
In general $\mathbb{P}(\Delta E|E) = \mathbb{P}(\Delta E|E, L)\cdot\mathbb{P}(L|E)$.
The form of $\mathbb{P}(L|E)$ is trivially available already from GEANT4 and it is easy to adapt the MC sampling to a vectorized computation on a GPU.
Contrarily, the form of $\mathbb{P}(\Delta E|E, L)$ is highly non-trivial to formulate.
Therefore, the scope of this study isolates the rigorous evaluation of the energy-loss components of the simulation.
While the generation of secondary particles is explicitly modeled, their subsequent propagation is temporarily disabled.
Likewise, the MSC is also disabled, since it is insignificant for this POC discussion.
This controlled environment allows for a precise validation of the main dynamics of the primary particles.
Extending the framework to propagate the generated secondaries and incorporate MSC represents a straightforward, structural evolution.

\section{The classical MC approach}\label{sec:the-classical-mc-sampling-approach}
\setlength{\parskip}{1em}
\setlength{\parindent}{0.5em}
Before deriving the unknown underlying straggling functions that govern the GEANT4 MC sampling, it is essential to first introduce the MC algorithm itself -
the modified-Urb\'an model~\cite{lassila1995energy, G4MAN} of energy-loss.
This model has been instrumental in representing, accurately and efficiently, the fluctuation in energy lost by charged particles propagating through matter, exciting and ionizing the target material.
While we fully document the model in App.~\ref{sec:appendix_a}, we concisely introduce its key features also here.

The primary motivation for employing this model lies in the breakdown of the `free electron' approximation within thin absorbers. 
In this regime, the energy-loss is often comparable to the atomic binding energies of the material, breaking the free electrons approximation.
The model addresses this by phenomenologically decomposing the energy-loss into two distinct physical processes: excitation and ionization.
Excitation accounts for the interaction with bound atomic shells at quantized energy levels approximated here as two fixed energy levels, satisfying $f_1 \ln{E_1} + f_2 \ln{E_2}=\ln{I}$, where $E_{1,2}$ are the energy levels, $f_{1,2}$ are the oscillator strengths ($f_1+f_2=1$) and $I$ is the mean ionization energy.
The current GEANT4 implementation considers only one excitation level by setting $f_2=0$ and hence $E_1=I$.
More generally, $E_2 = 10~{\rm eV} Z^2$ and $f_2=0$ if $Z=1$ or $2/Z$ if $Z\geq 2$, where $Z$ is the medium atomic number.
Ionization captures the stochastic, continuous spectrum of higher energy transfers governed by the Rutherford cross-section that behaves as $\frac{1}{E^2}$.

\begin{figure}[t]
\centering    \includegraphics[width=\linewidth]{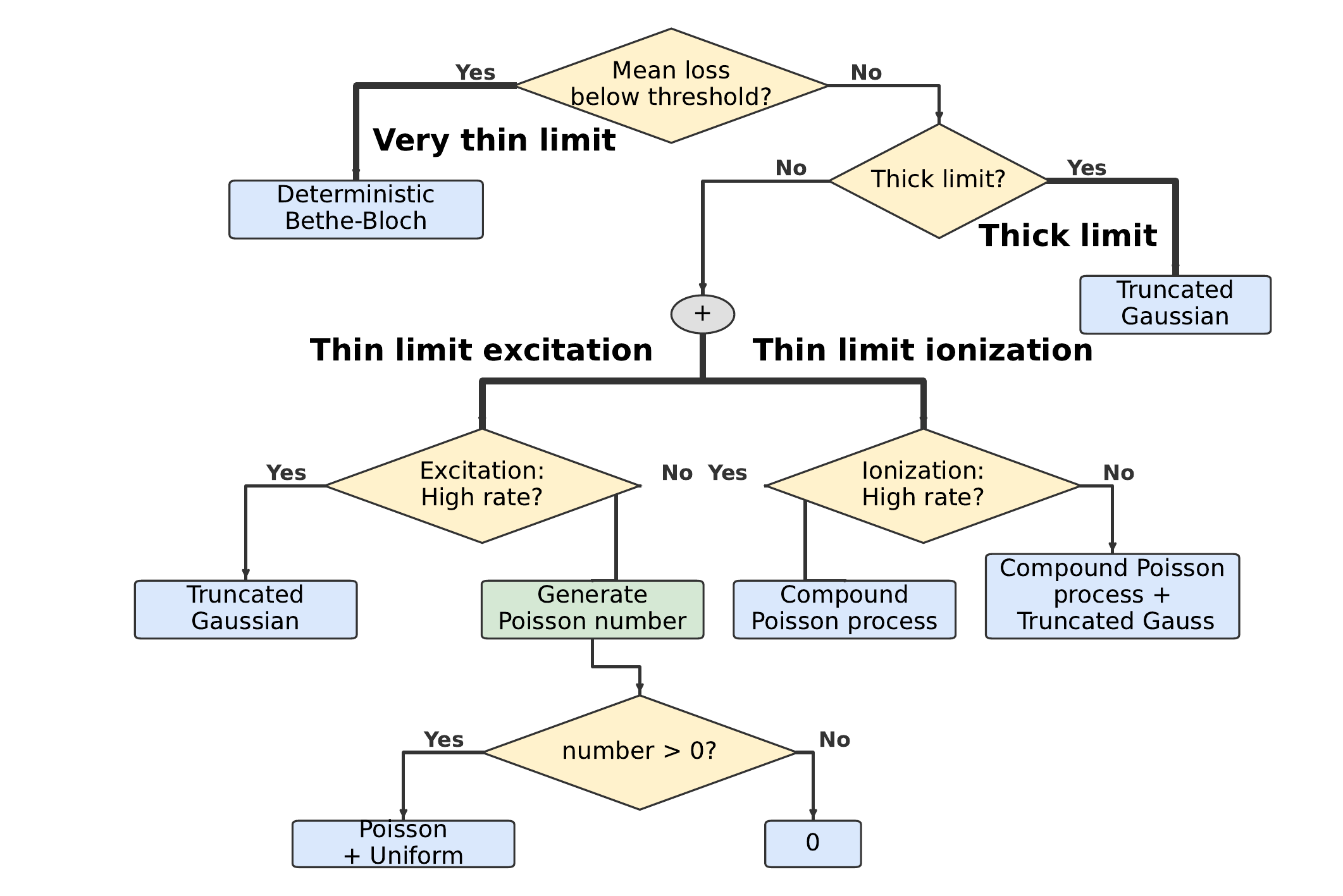}
\caption{The MC decision tree for energy-loss through continuous excitation and ionization.
All decisions are made based on the kinetic energy of the primary particle and the step-length, leading to a few specific distributions from which the energy-loss is sampled.
The ``+'' sign represents addition of the energy-loss sampled from both of the branches.}
\label{fig:cnt_mc_decision_tree}
\end{figure}

To mitigate the computational cost of a detailed collision-by-collision simulation, the model implements a regime-dependent approximation strategy. 
It implicitly partitions the $E\textsf{--}L$ phase-space based on the collision frequency (average number of collisions per step).
At low collision rates, it retains the full rigor of the exact models, whereas at high rates, it leverages the asymptotic behavior described by the Central Limit Theorem (CLT) to model parts (or all) of the energy-loss fluctuations as a Gaussian process.
Furthermore, a production-energy threshold, $T_{\rm cut}$, is assumed for classifying ionization collisions.
This threshold is defined in GEANT4 only by the medium and the length-precision set by the user per particle~\cite{agostinelli2003geant4, 1610988, ALLISON2016186}.
For example, for protons in aluminum, with the range-cut set to 1~$\mu{\rm m}$ (leading to a very high granular simulation), it is $T_{\rm cut}=1010.57$~eV.
Energy transfers below this threshold are accumulated into the continuous energy-loss of the step, while transfers exceeding it result in the explicit generation of secondary electrons and the termination of the step.
The continuous energy-loss via excitation collisions accumulates alongside ionization independently of $T_{\rm cut}$.
The decision tree of the continuous energy-loss sampling algorithm is presented in Fig.~\ref{fig:cnt_mc_decision_tree}.
In the following few paragraphs, we navigate through the main four branches of it.
The full derivation of the different terms in it is given in App.~\ref{sec:appendix_a}.
Hereafter, the `ion', `exc', `cnt' and `sec' notations stand for ionization, excitation, continuous energy-loss and energy-loss via secondary production for simplicity.

\noindent\textbf{Very thin limit:} Initially, it is checked if the mean energy-loss, $\langle \Delta E\rangle$, is below a fixed threshold of $10\text{ eV}$. If this condition is met, the energy-loss is deterministically fixed to the mean loss, given by the Bethe-Bloch formula~\cite{ParticleDataGroup:2024cfk} of the stopping power, $\langle\frac{dE}{dx}\rangle$, where $\langle \Delta E\rangle = \langle\frac{dE}{dx}\rangle \cdot L$.
In practice, $\langle\frac{dE}{dx}\rangle$ pre-calculated by GEANT4 during the initialization phase and stored in a fine-grained lookup-table conditioned on $E$.

\noindent\textbf{Thick limit:} This limit is typical for regions in the phase-space, where high granularity is not required and it is explicitly defined by:
$\left<\Delta E\right> > \kappa T_{\rm cut}$,
where $\kappa$ is fixed to 10.
In this regime, the particle undergoes a sufficiently large number of collisions, between generation of secondary electrons, such that the statistical fluctuations are dominated by the CLT.
Additionally, the atomic structure (binding energies) is neglected, allowing the scattering process to be modeled as interactions with free electrons.
Consequently, the straggling function in this regime converges to a Gaussian distribution with $\langle \Delta E\rangle$ as the mean energy-loss and Bohr's formula for free electrons as the variance, $\Omega^2$.

\noindent\textbf{Thin limit ionization:}
The most accurate realization of energy-loss through a single ionization collision  is:
\begin{equation}
\label{eq:e_ion_colision}
\Delta \mathbb{E}_{\rm ion}^{\rm coll.} = \frac{E_0}{1 - u\frac{T_{\rm max} - E_0}{T_{\rm max}}}\text{,}
\end{equation}
where $E_0$ is the minimum energy-loss, $T_{\rm max}$ is the maximum energy transfer in a single collision and $u$ is a random variable sampled from a uniform distribution in [$0, 1$].
As the mean number of collisions in a single step increases, the simulation becomes computationally prohibitive.
To balance accuracy with efficiency, the modified-Urb\'an model introduces a dynamic energy threshold, $w_3$, proportional to $E_0$.
Below this threshold the collisions are considered `soft' and therefore aggregated into a Gaussian term.
The few `hard' collisions above $w_3$, but yet bellow $T_{\rm cut}$, are explicitly resolved as individual collisions to deal with the asymmetric `Landau tail' of the straggling function.
Consequently, the continuous ionization is sampled as following:
\begin{equation}
\label{eq:continuous_ionization}
\Delta E_{\rm ion}^{\rm cnt} =
\begin{cases}
\sum_{j=1}^{{\rm Poisson}\left(\langle n_3 \rangle \right)} \frac{E_0}{1 - u_j\frac{T_{\rm cut} - E_0}{T_{\rm cut}}}, \\
\quad \text{if } \langle n_3 \rangle < n_{\max}^{\mathrm{cnt}} \\[1.5em]
\sum_{j=1}^{{\rm Poisson}\left(\langle n_3 \rangle - \langle n_a \rangle \right)} \frac{w_3}{1 - u_j\frac{T_{\rm cut} - w_3}{T_{\rm cut}}} \\
+{\rm Gaussian} \left( \langle \Delta E_{\rm ion}^{\rm soft} \rangle, \sigma^{\rm soft} \right) \\
\quad \text{otherwise,}
\end{cases}
\end{equation}
where $\langle n_3 \rangle$ is the mean number of ionization collisions, $n_{\max}^{\rm cnt}$ is the (maximum) threshold number of collisions.
The term $\langle n_a \rangle$ is the number of `soft' collisions tackled by the Gaussian sampling, that on itself is characterized by a mean of $\langle \Delta E_{\rm ion}^{\rm soft} \rangle$ and width of $\sigma^{\rm soft}$.
The proper definition of all terms is given in App.~\ref{sec:appendix_a}.
This kind of summation, where the summand is a random variable and the number of summed variables are sampled from a Poisson distribution is recognized as a `compound Poisson process'~\cite{Iacus2018}.

\noindent\textbf{Thin limit excitation:} When the mean number of excitation collisions exceeds $n_{\max}^{\rm exc}$ (a threshold fixed to 8), the energy-loss is approximated to a Gaussian number (via the CLT) sampled using a mean $\langle \Delta E_{\rm exc} \rangle$ and width $\sigma_{\rm exc}$.
The proper definition of all terms is given in App.~\ref{sec:appendix_a}.
Values sampled outside $[0,2\langle \Delta E_{\rm exc} \rangle]$ are resampled until falling in that range.
When the mean excitation collisions is below $n_{\max}^{\rm exc}$, the energy-loss is sampled more rigorously, from a complex mixture of Poisson and uniform distributions,
\begin{equation}
\label{eq:e_exc_sampling}
\Delta E_{\rm exc}^{\rm cnt} = (\text{Poisson}(\lambda_{\rm exc}) + 1 - 2 \cdot \text{Uniform}(0,1))E_1\text{,}
\end{equation}
where $\lambda_{\rm exc}$ is the mean number of excitation collisions.

\begin{figure}[!ht]
\centering
\includegraphics[width=\linewidth]{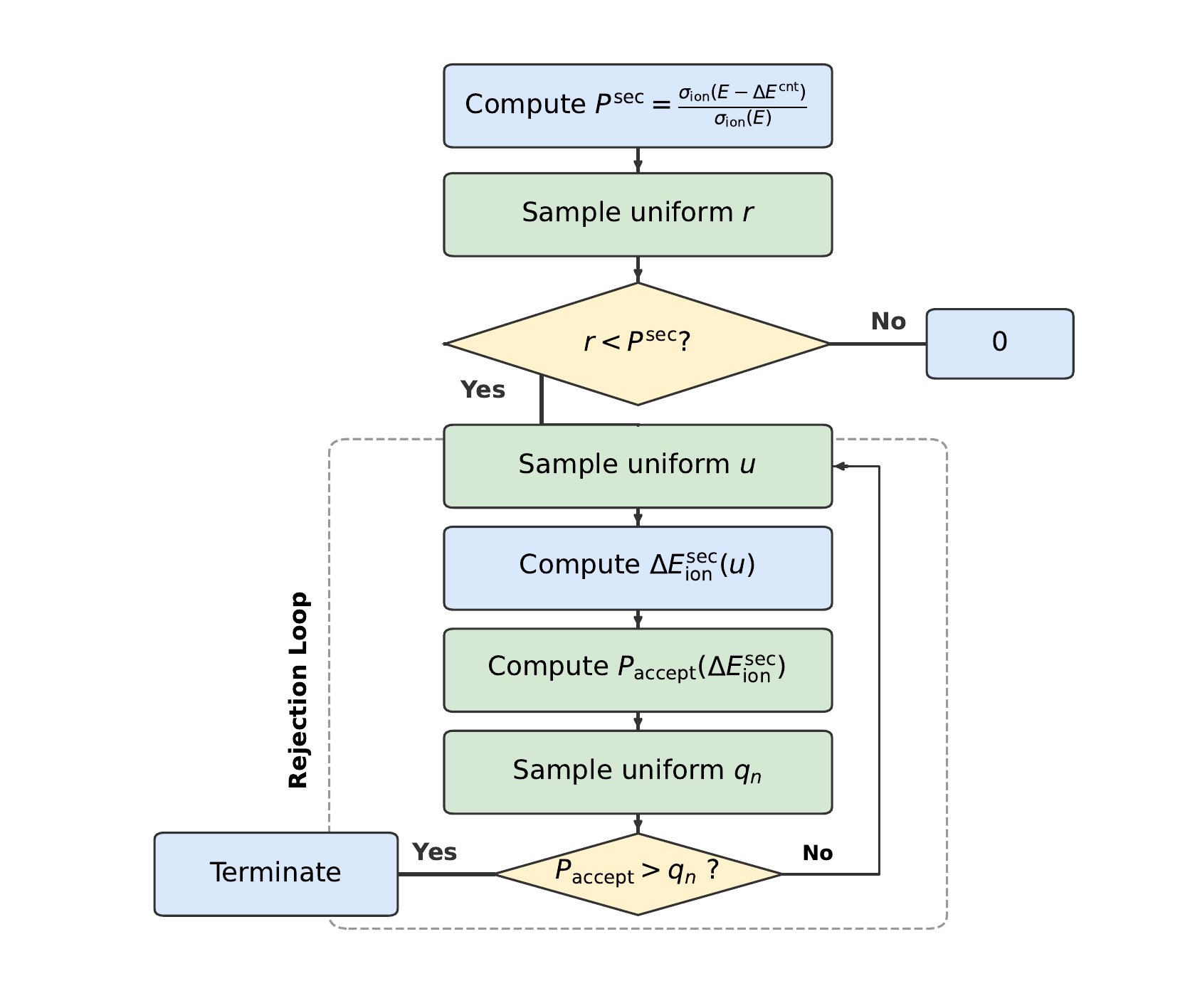}
\caption{The GEANT4 MC decision tree for energy-loss through generation of a secondary electron.}
\label{fig:sec_mc_decision_tree}
\end{figure}

\noindent\textbf{Secondary electron generation:} 
Finally, when a secondary electron is produced through a discrete ionization (at the end of the step), an additional fluctuation term is added to the energy loss.
The probability of this discrete emission relies on a fine-grained lookup table evaluated as a function of the primary particle's energy, $E$.
This lookup table is pre-calculated by GEANT4 during the initialization phase using deterministic theoretical cross-sections for ionization, $\sigma_{\rm ion}$, rather than MC methods ($\sigma_{\rm ion}$ also dictates the step-length physics PDF: $\mathbb{P}(L|E)\propto e^{-\sigma_{\rm ion} L}$).
Generally, $\sigma_{\rm ion}$ depends on $E$ and $T_{\rm cut}$.
The energy of the secondary electron is sampled with the same form of $\Delta \mathbb{E}_{\rm ion}^{\rm coll.}$ after passing the rejection loop presented in the decision tree of Fig.~\ref{fig:sec_mc_decision_tree} to account for the full differential cross-section.
More details are given in App.~\ref{sec:appendix_a}.

\section{The effective straggling function}\label{sec:the-effective-straggling-function}
\setlength{\parskip}{1em}
\setlength{\parindent}{0.5em}
Since the Gaussian limits are straightforward, we restrict the following discussion to the non-trivial derivation of PDFs for the low collision rate regime and the secondary energy rejection strategy.

The full derivation is detailed in App.~\ref{sec:appendix_b}, and the final PDFs, $\mathbb{P}_{\Delta E}(x)$, where $x$ denotes the specific energy-loss, are summarized in Eq.~\ref{eq:e_exc_pdf}--\ref{eq:e_i_ion_full_characteristic_expression}.
These PDFs are derived here for the first time.
They are completely generic and can be widely used to estimate the total energy-loss at any point in the $E\textsf{--}L$ phase-space.
In the reminder of this study, we use them during the training of our generative model as discussed in Sec.~\ref{sec:physics-informed-generative-model}.

The continuous excitation PDF is given by
\begin{equation}
\label{eq:e_exc_pdf}
\mathbb{P}_{\Delta E_{\rm exc}^{\rm cnt}}(x) = \frac{e^{-\lambda_{\rm exc}}}{2}\left(\frac{\lambda_{\rm exc}^{\lfloor x \rfloor}}{\lfloor x \rfloor!} + \frac{\lambda_{\rm exc}^{\lfloor x \rfloor + 1}}{(\lfloor x \rfloor + 1)!}\right).
\end{equation}

The PDF of a compound Poisson process is given by integral formula:
\begin{equation}
\label{eq:e_ion_pdf_integral}
\mathbb{P}_{\Delta E^{\rm cnt}_{\rm ion}}(x) = \frac{1}{2\pi} \int_{-\infty}^{\infty} e^{-\mathbbm{i}xt} e^{\lambda_{\rm ion} \left( \phi(t) - 1 \right)} \, {\rm d}t,
\end{equation}
where $\phi(t)$ denotes the characteristic function of the summand and $\lambda_{\rm ion}$ is the Poisson argument from either term in Eq.~\ref{eq:continuous_ionization}.
The explicit form of the single-collision characteristic function, as derived in App.~\ref{sec:appendix_b}, is:
\begin{align}
\label{eq:e_i_ion_full_characteristic_expression}
\phi_{\Delta \mathbb{E}_{\rm ion}^{\rm coll.}}(t) &= \frac{\cos(\omega t)}{\omega} + \mathbbm{i}\frac{\sin(\omega t)}{\omega}  \notag \\
& - \frac{\cos\left(\rho t\right)}{\rho} - \mathbbm{i}\frac{\sin\left(\rho t\right)}{\rho}  \notag \\
& + t\mathrm{Si}(\omega t) - \mathbbm{i}t\mathrm{Ci}(\omega t)  \notag \\
& - t\mathrm{Si}\left(\rho t\right) + \mathbbm{i} t\mathrm{Ci}\left(\rho t\right)\text{,}
\end{align}
where $\mathrm{Si}$ and $\mathrm{Ci}$ are the standard sine and cosine integral functions.
In the original Urb\'an model $\omega=E_0$ and $\rho=T_{\rm max}-I$.
In the GEANT4 implementation, $\omega=w_3$ and $\rho=T_{\rm cut}$.
The strong oscillatory nature of $\phi_{\Delta \mathbb{E}_{\rm ion}^{\rm coll.}}(t)$ implies that the integral in Eq.~\ref{eq:e_ion_pdf_integral} has no trivial analytical solution, when substituting $\phi(t)$ with $\phi_{\Delta \mathbb{E}_{\rm ion}^{\rm coll.}}(t)$.
Further, explicit numerical integration attempts are deemed not parctical.
However, we identify that it is simply an inverse Fourier transform and can be evaluated precisely using its discrete form.

The PDF describing the discrete ionization leading to the generation of secondary electrons is given by
\begin{equation}
\label{eq:discrete_ionization_pdf}
\mathbb{P}_{\Delta E_{\rm ion}^{\rm sec}}(x) = \frac{T_{\max} - \beta^2x}{\left(\frac{T_{\max}}{T_{\rm cut}} - 1 - \beta^2 \ln\left(\frac{T_{\max}}{T_{\rm cut}}\right)\right)x^2}\text{,}
\end{equation}
where $\beta$ is the particle relativistic speed.

Finally, while in the classical MC sampling approach, the total energy-loss is simply the sum of the different contributions, the PDF description of the total loss is rather obtained from the convolution of the individual PDFs.
This result is the ``effective straggling function''.

\begin{figure*}[!ht]
\centering
\includegraphics[width=\linewidth]{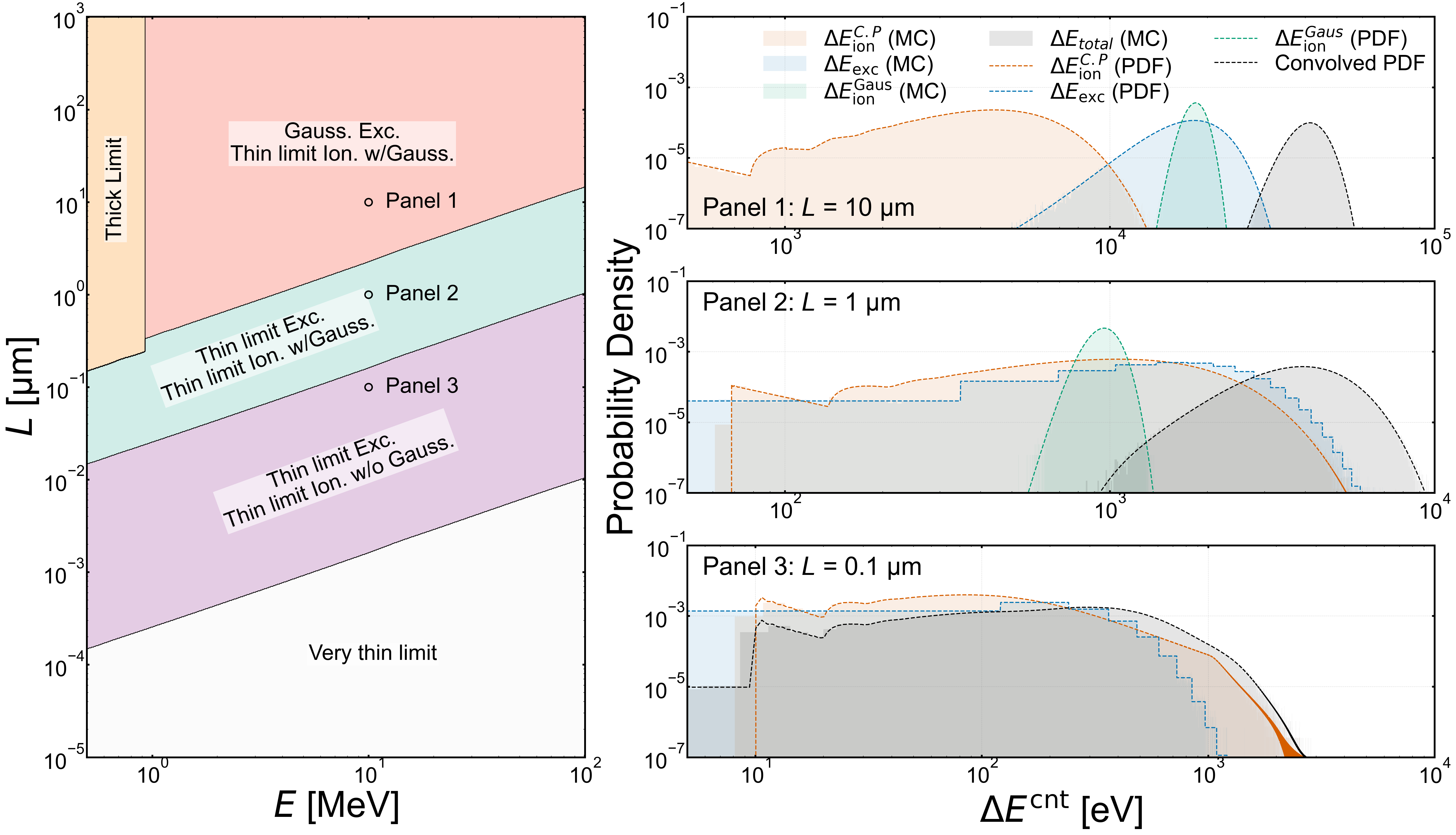}
\caption{The implicit division of the $E\textsf{--}L$ phase-space (for a range-cut of $1 \mu \rm m$) as implemented in GEANT4 is shown on the left. The MC distributions of the energy-loss compositions discussed in Sec.~\ref{sec:the-classical-mc-sampling-approach} are plotted on the right for three example points (marked on the left) along with the respective PDFs.}
\label{fig:pdf_phase_space}
\end{figure*}

Examples of the individual PDFs as well as their convolution, at different points in the $E\textsf{--}L$ phase-space are illustrated in Fig.~\ref{fig:pdf_phase_space}.

The MC behavior is perfectly captured by these PDFs.
It is particularly possible to see, for the first time, how complex these PDFs can grow, having different sharp features, spanning huge range of energy-loss values, etc.
We have checked that the MC shapes are perfectly captured in the full phase-space of this study.
In terms of computational performance, the evaluation of a PDF takes at most $\mathcal{O}(1~{\rm s})$ in the most complex regions, and significantly less in others.
It is important to note that this benchmark reflects a non-optimized Python implementation; a dedicated C++ integration would naturally yield substantially faster execution times.
In the context of this study, the explicit calculation of all PDFs needs to be done only once, before the training of the ML model.
We now turn to describe the ML framework and how the physics PDFs are incorporated into it.

\section{The training dataset}\label{sec:the-training-dataset}
\begin{figure}[!ht]
\centering
\includegraphics[width=0.45\textwidth]{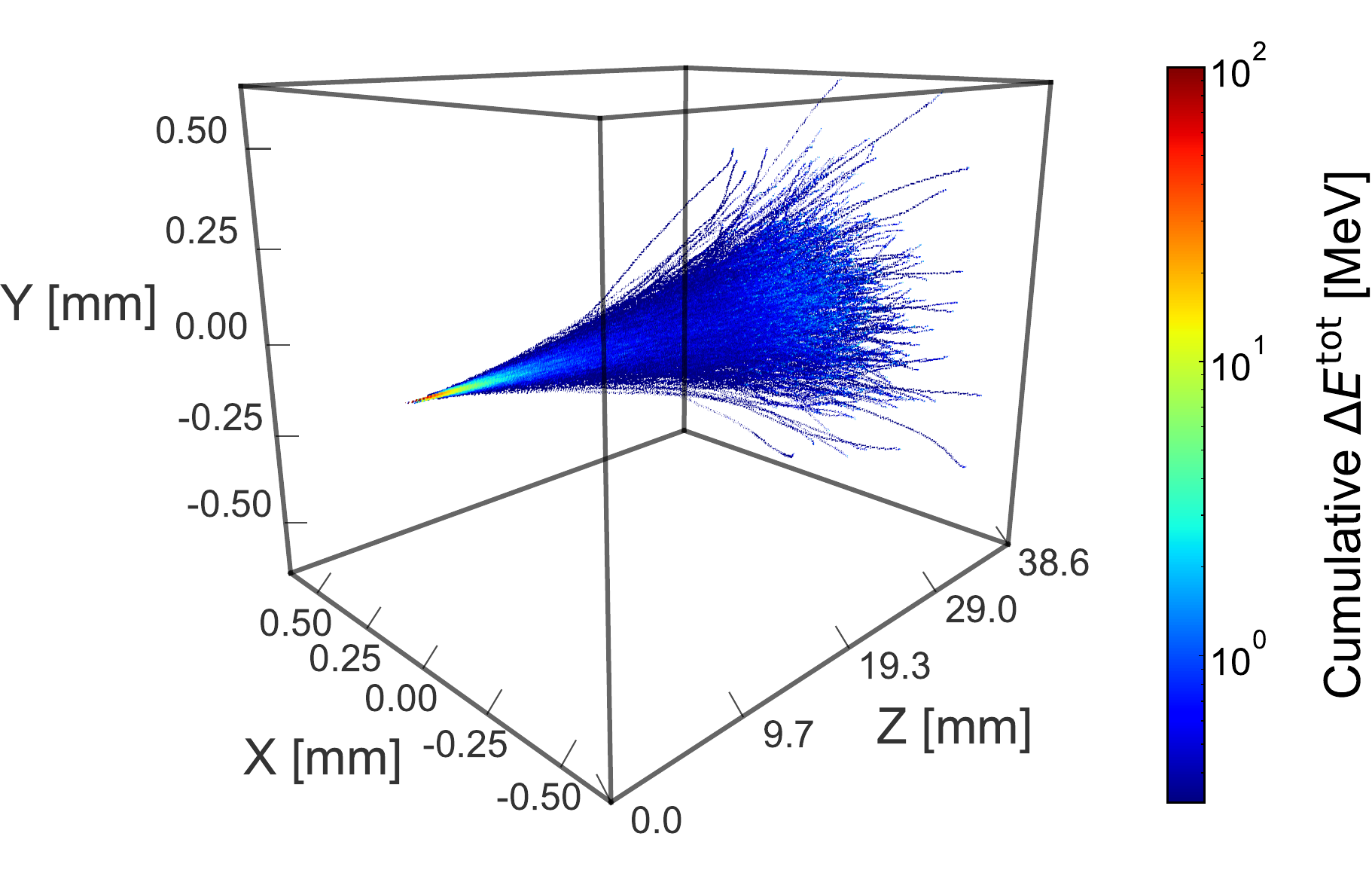}
\caption{Cumulative energy loss by 1,000 primary protons trajectories of 100~MeV each, propagating through an aluminum bulk until full stop. The 3D histogram is binned in to $250\times250\times1000$ slices.}    \label{fig:geant4_tracks}
\hfill
\end{figure}

\setlength{\parskip}{1em}
\setlength{\parindent}{0.5em}
We use GEANT4 to generate a few data samples for the training.
Each sample contains 1,000 primary protons shot with $100~\text{MeV}$ on an aluminum bulk that is long enough to absorb all primary protons.
Using a range-cut of $1~\mu{\rm m}$, each proton yields $\sim$10,000 steps.
The protons' trajectories are presented in figure~\ref{fig:geant4_tracks}.
The range-cut is set here to a specifically very small value in order to capture the full fidelity of the propagation and energy-loss profiles.
Typically, simulations of thick passive materials (e.g. for calorimeters and dead shielding involving high-$Z$ materials) are using much larger range-cut at the level of $1\textsf{--}10$~mm (see e.g.~\cite{ATLAS:2010arf}) to save CPU time. 
Applying such macroscopic range-cuts to our setup would reduce the number of steps per trajectory to only a few, entirely obscuring the fine-grained spatial energy deposition profile.

For each primary proton, we record the following features per step: the kinetic energy at the beginning of a step, the position and momentum direction before and after the step, the continuous energy-loss along the step, the energy-loss associated with the production of secondary electrons during the step, the number of secondary electrons produced along the step and the name of the physical process, which determined the step.
The time feature is computed analytically from the energy at the beginning of the step and step length and hence is omitted from this list.
In this study we focus on the modeling of propagation of the primary particles.
The initial conditions needed to generate the secondary electrons are explicitly computed for every step, however, we ignore their propagation since the associated mechanism is similar to the one used for the primary particles.
We leave this development for a future work.

Once a simulation is initialized, the governing physics at a specific step is determined based on the particle's identity, its kinetic energy and the medium solely.
The value of $L$ is obtained from the GPU-accelerated MC algorithm.
Training a generative network to simulate the energy-loss through continuous ionization and excitation $\Delta E^{\rm cnt}_{\rm ion}+\Delta E^{\rm cnt}_{\rm exc}$, secondary electrons via ionization $\Delta E^{\rm sec}_{\rm ion}$ and the deflection polar angle $\theta\in[0, \pi]$, is sufficient to recapture GEANT4 computations.
Within a given step, the features are computed in the following sequence, given $E$: $L \longrightarrow \Delta E^{\rm cnt} \longrightarrow \Delta E^{\rm sec}, \theta$.
The azimuthal deflection angle $\varphi$ is redundant for the training since its distribution is uniform in [0, $2\pi$].
Finally, when MSC is turned off, the position at the end of the step is determined from $\theta$, $\varphi$ of the previous step and $L$, while the step time is determined analytically from $E$ and $L$.

\begin{figure*}[!ht]
\centering
\includegraphics[width=\textwidth]{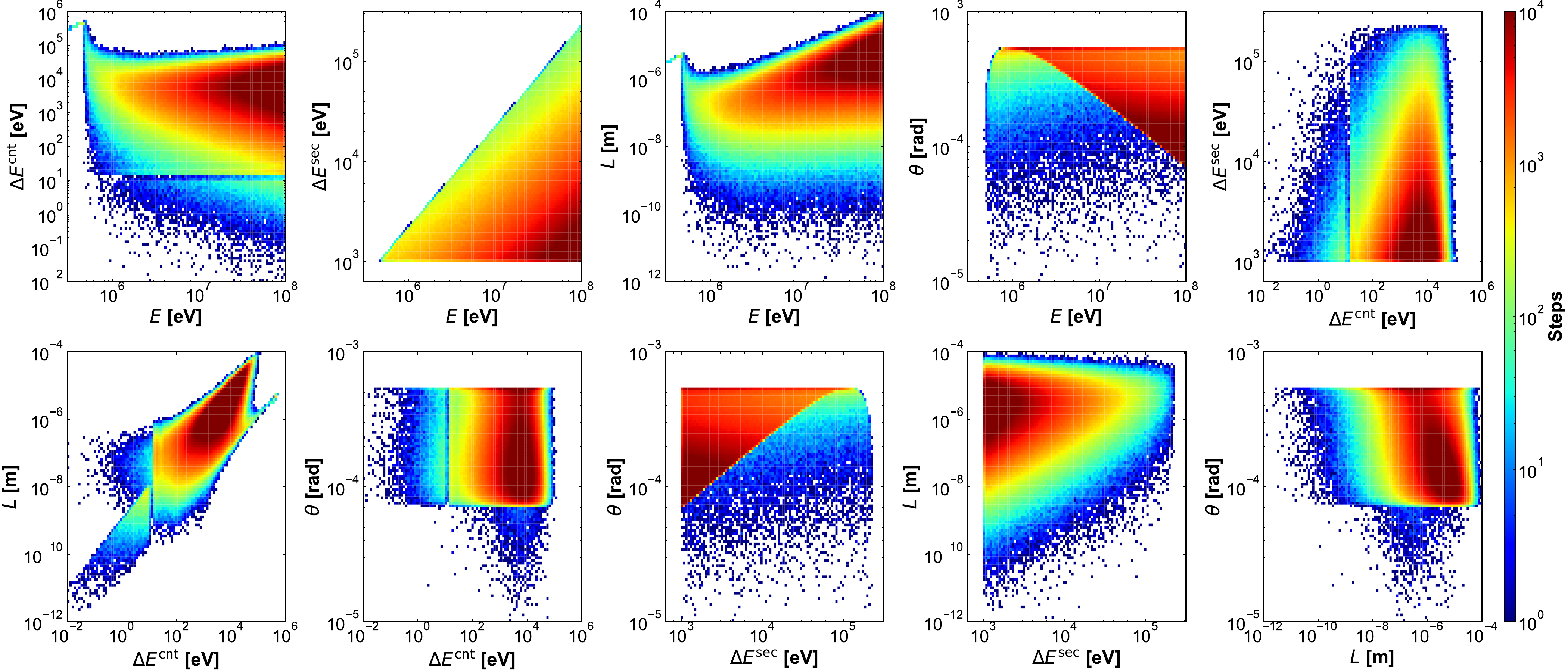}
\caption{The relations between all features (kinetic energy, step length, deflection angle, continuous and secondary generation energy-loss) of one thousand 100~MeV protons propagating through aluminum and depositing energy until stopped and absorbed. The $z$ axis scale corresponds to the number of steps.}
\label{fig:cross_features}
\hfill
\end{figure*}

The pair-wise distributions of the training features are presented in figure~\ref{fig:cross_features}.
The ML framework task is to learn these relations.
Already from this picture it can be easily understood that this is a highly non-trivial task to execute, even in the limit of an infinite dataset.
Hence, to improve the learning capability of the model, we introduce the underlying physics information outlined above in Sec.~\ref{sec:the-effective-straggling-function}, as discussed below.

\section{Physics informed generative model}\label{sec:physics-informed-generative-model}
\setlength{\parskip}{1em}
\setlength{\parindent}{0.5em}
In this section, we first briefly introduce and justify the choice of the baseline ML model used in this work.
We later discuss how the physics information is incorporated into its PHIN variant.

The prediction of PDFs for non-deterministic problems in physics is often addressed through computationally intensive MC simulations.
Among DGMS, Variational Autoencoders (VAEs), Normalizing Flows, Diffusion models, and GANs represent the primary architectural paradigms~\cite{Hashemi_2024}.
While VAEs offer fast generation, they often yield over-smoothed distributions that struggle to capture the sharp kinematic boundaries required in fundamental physics.
More recently, Normalizing Flows and Diffusion models have emerged as the state-of-the-art for exact likelihood estimation and high-fidelity sample generation, respectively.
However, their reliance on complex invertible transformations or iterative denoising steps incurs a high computational cost during inference.
Because the primary objective of surrogate modeling in this context is simulation acceleration, this inference latency presents a prohibitive bottleneck. 
GANs, conversely, execute generation in a single forward pass, providing a critical advantage in execution speed.

Standard GANs employ a distinctive two-network structure--a generator producing candidate samples and a discriminator evaluating them--trained in a zero-sum game.
However, their learning procedure is notoriously unstable and often leads to mode collapse~\cite{arjovsky2017wasserstein}, which occurs when the generator produces a limited variety of outputs, ignoring diverse patterns in the training data.
Furthermore, standard GAN loss curves lack interpretable correlation with sample quality, making it challenging to optimize divergence metrics and hyperparameters.
To address these key issues, \textcite{arjovsky2017wasserstein} introduced the Wasserstein~GAN (WGAN).
In a WGAN, the discriminator is replaced with a critic function, trained to maximize the difference between its prediction on real and generated data using the following loss function:
\begin{equation}
\label{eq:wgan_loss}
\mathcal{L}_{\rm Wass.}^{\rm data} = \mathbb{E}_{\mathbf{x} \sim p_{\text{data}}} [D(\mathbf{x})] - \mathbb{E}_{\mathbf{z} \sim p_{\mathbf{z}}} [D(G(\mathbf{z}))] \text{, }
\end{equation}
where $\mathbb{E}$ denotes the expected value, $D$ is the critic function, $G$ is the generator, $\mathbf{x}$ is the real data, and $\mathbf{z}$ is the noise input to the generator.
To further enhance stability, \textcite{gulrajani2017improved} proposed penalizing the norm of the critic's gradient with respect to its input, replacing standard weight clipping.
This gradient penalty (GP) significantly improves WGAN convergence and stability.
This improved formulation, WGAN-GP, serves as the baseline ML architecture in this work.
Despite these advancements, training such networks remains inherently challenging.
Minimizing the loss function does not guarantee convergence to a physically meaningful solution, and instability may still arise.
In this context, the introduction of domain-specific knowledge -- particularly physics-informed constraints through well-defined loss functions -- offers a compelling strategy to guide the learning process, improving both convergence behavior and the physical fidelity of the generated outputs.

A crucial requirement for this application is the ability to condition the generation of simulation steps on the incident particle's dynamic state.
Therefore, we integrate the concept of conditional GANs~\cite{mirza2014conditional}, which allows the model to generate data following specific conditions by incorporating this information into both the generator and critic networks.
Hereafter, we refer to this baseline conditional WGAN-GP simply as `GAN', and its physics-informed variant as `PHIN-GAN'.

The analytical PDFs formulated in Eq.~\ref{eq:e_exc_pdf}--\ref{eq:discrete_ionization_pdf} are incorporated in two novel fashions: (i) precisely penalizing the generator throughout the training process, and (ii) unique physics-based scaling of the datasets. These two concepts are detailed below.

\noindent\textbf{Physics penalization:} the technique in which we incorporate the underlying physics as a distribution-level learning objective, by penalizing the generator throughout the training session for producing samples that deviate from the target PDFs.
This is realized by the Kolmogorov-Smirnov (KS) distance, which is a non-parametric test that quantifies the probability that two sets of 1D values originate from the same underlying distribution.
Since the PDFs of the continuous and secondaries energy-losses depend on the particle state, a standard KS computation (which handles unconditional distributions) is not sufficient.
To handle this problem, we first mark a grid of 2D (1D) points logarithmically spaced in the $E\textsf{--}L$ ($E$) phase-space.
We have found that it is enough to use $N_{\rm pts}=1265$ (200) points, where the continuous (secondaries) energy-loss PDFs are pre-computed.
This computation is done once per point before the training.
Then, during the training, we generate 10,000 steps per point with the PHIN-GAN and compute the KS distance, $\partial^{\rm cnt}_{\rm KS}$ and $\partial^{\rm sec}_{\rm KS}$, between the distribution of the generated energy-loss values and the corresponding physics PDF per point.
The two physics-loss functions of the PHIN-GAN are computed as the average KS distance over all points in the phase-space for either the continuous or the secondaries PDFs:
\begin{equation}
\label{eq:physics_loss}
\mathcal{L}_{\rm phys}^{\rm prc} = \frac{\alpha^{\rm prc}}{N_{\rm pts}^{\rm prc}}\sum_{j=1}^{N_{\rm pts}^{\rm prc}} (\partial_{\rm KS}^{\rm prc})_j\text{,}
\end{equation}
where ${\rm prc}$ (process) is $\{{\rm cnt,sec}\}$ and where $\alpha$ are two arbitrary weights.
These values are determined empirically and the dependence of the results on their specific choices is not large as long as they are kept within a reasonable range.
Specifically, small weights have small impact on the training process, while excessively high ones may lead to over-regularization that prevents the network from learning other important features.

\noindent\textbf{Physics-based data scaling:} the simplest and most widely used scaling methods are: \emph{standardization} and \emph{normalization}. These are typically defined as $x \mapsto \frac{x - \mu}{\sigma}$ and $x \mapsto \frac{x - x_{\rm min}}{x_{\rm max} - x_{\rm min}}$, respectively, where $\mu$, $\sigma$, $x_{\rm min}$, $x_{\rm max}$ are the global mean, standard deviation, minimum and maximum value of the dataset.
In many cases, however, these global transformations may be too inclusive to capture both large and small features of the distribution at once.
This is particularly true for cases in which the data may span many orders of magnitude and/or may have non-trivial behavior (sharp features, etc.), as in ours.
Furthermore, these transformations are subjected to the statistics of the training dataset.
The training dataset of the GAN is scaled with the global transformations mentioned above.
For the PHIN-GAN, instead of blindly applying global standardization or normalization based on the data, we exploit our prior physics knowledge of the system.
Specifically, the distribution of $\Delta E^{\rm sec}$ has well-defined lower and upper bounds, $T_{\rm cut}$ and $T_{\rm max}$, both of which depend only on $E$.
This allows us to perform a \emph{functional normalization} that and can be tailored precisely at any point in the phase-space.
Similarly, the $\Delta E^{\rm cnt}$ distribution depends on both $E$ and $L$ and its analytical form is known. 
Therefore, the local mean and standard deviation can be precomputed on an arbitrarily fine grid in the $E\textsf{--}L$ space (we use $5000\times 5000$), enabling \emph{local standardization}.
We stress that this kind of a transformation, which exploits the physics information, is practically impossible in case the learning is done from data alone.
The remaining features ($E$, $L$ and $\theta$) undergo logarithmic scaling irrespective of the physics knowledge, as described in App.~\ref{sec:appendix_d}.

\begin{figure*}[!ht]
\centering
\includegraphics[width=\textwidth]{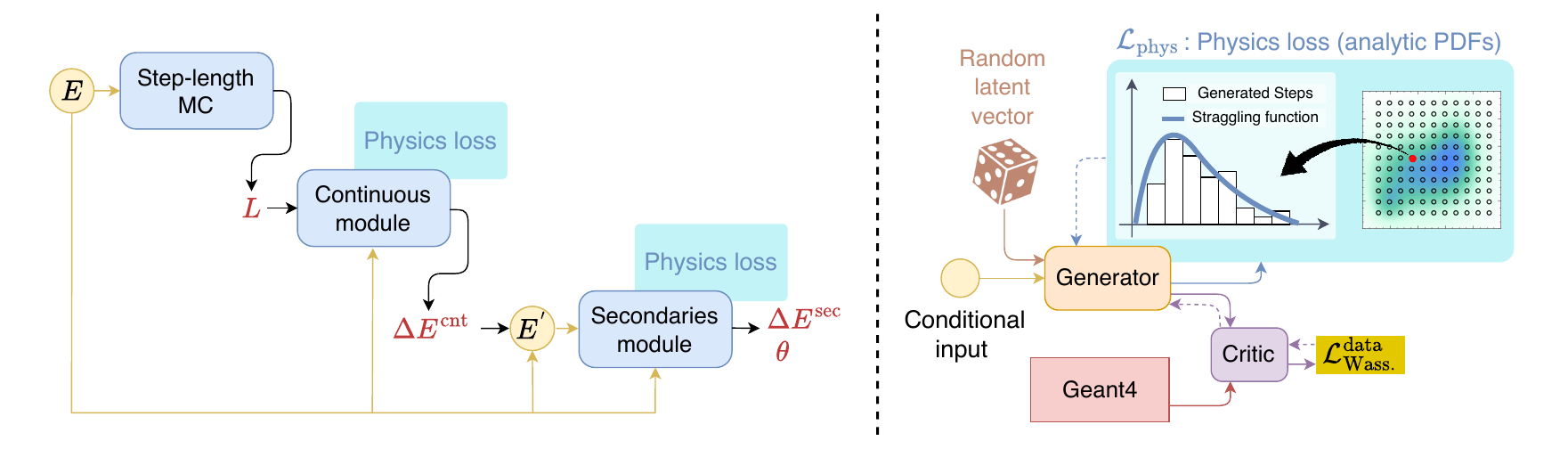}
\caption{
Schematic overview of PHIN-GAN's architecture.
For each step, PHIN-GAN computes the features in a sequential hierarchy identical to the GEANT4 stepping logic.
This is done by the GPU-accelerated step-length MC algorithm and the continuous and secondaries modules, each of which contains its own generator and critic.
Given $E$, first $L$ is generated, then $\Delta E^{\rm cnt}$ is generated based on $E$ and $L$, then $\Delta E^{\rm sec}$ and $\theta$ are generated based on $E'=E-\Delta E^{\rm cnt}$.
The conditioning is performed with an embedding layer, to which a random latent vector is concatenated to allow the stochastic nature of the step.
Crucially, the generators are regularized by the physics block using the energy-loss PDFs from Sec.~\ref{sec:the-effective-straggling-function}.
Each module is trained independently, where during the training of the secondaries module, the values of $E'$ from GEANT4 are used.
}\label{fig:gan_architecture}
\hfill
\end{figure*}

The PHIN-GAN architecture, illustrated in Fig.~\ref{fig:gan_architecture}, mimics the three-stage simulation sequence of GEANT4.
First, a step length ($L$) is generated, conditioned solely on the incident kinetic energy ($E$).
Next, the continuous energy loss, $\Delta E^{\rm cnt}$, is generated conditioned on both $E$ and $L$. 
Finally, the secondaries energy-loss ($\Delta E^{\rm sec}$) and the scattering angle ($\theta$) are generated conditioned on the remaining energy, $E'=E-\Delta E^{\rm cnt}$.
The first stage is executed as a GPU-accelerated MC as mentioned above, whereas the latter two stages employ independent \textit{generators}.
Each \textit{generator} is trained adversarially against a dedicated \textit{critic}.
Detailed network specifications are provided in Fig.~\ref{fig:networks_architecture}.

\section{The Training Setup}\label{sec:the-training-setup}
\setlength{\parskip}{1em}
\setlength{\parindent}{0.5em}
With the data scaled and the outlined network architectures in place, we train both the GAN and the PHIN-GAN on a dataset of 1,000 trajectories of 100~MeV protons propagating through an aluminum bulk to a full stop. This dataset comprises $\sim 9.8\times 10^6$ steps.
Steps with $\Delta E^{\rm cnt}=0$, along with steps falling in the deterministic region of the $E\textsf{--}L$ phase-space  are excluded from the training dataset of the continuous module (\textit{generator} and \textit{critic}). 
These amount to 1.2\% of the all steps.
The energy-loss of steps in the very thin (deterministic) region is simply given by the mean loss described by the Bethe-Bloch formula.
Likewise, steps with no secondary electron generation are excluded from the training dataset of the secondaries module.
These amount to 18.4\% of the all steps.
All removed steps are re-introduced during the stepping inference.
For training, the samples are split into 64 batches.
The learning rate is set to $10^{-5}$ using an `RMSprop' optimizer~\cite{1370017282431050757,arjovsky2017wasserstein} for both modules.
We set the training frequency to 3:1, such that the {\it generator} undergoes back-propagation once for every three training steps of the {\it critic}, ensuring the {\it critic} remains further trained than the {\it generator}.
We add the physics-loss terms from Eq.~\ref{eq:physics_loss} to the loss function of the continuous and the secondaries generators using weights of $\alpha^{\rm cnt}=\alpha^{\rm sec}=0.1$.
The values of these weights were determined empirically and can, in principle, be further optimized.

We compute the physical loss terms, $\mathcal{L}_{\rm phys}^{\rm cnt}$ and $\mathcal{L}_{\rm phys}^{\rm sec}$ (defined by Eq.~\ref{eq:physics_loss}), throughout the training of both architectures, although they only serve as active penalties for PHIN-GAN.
As shown in Fig.~\ref{fig:losses_vs_epochs}, PHIN-GAN achieves significantly lower loss values and a more stably decreasing trend.
This demonstrates the stabilizing effect of the physics constraints on the learning dynamics.
By forcing the \textit{generator} to strictly adhere to the underlying physics, these constraints enhance sample fidelity and mitigate the overfitting behavior evident in the standard GAN's rising penalty trend.

\section{Results}\label{sec:results}
\begin{figure}[!ht]
\centering    \includegraphics[width=0.45\textwidth]{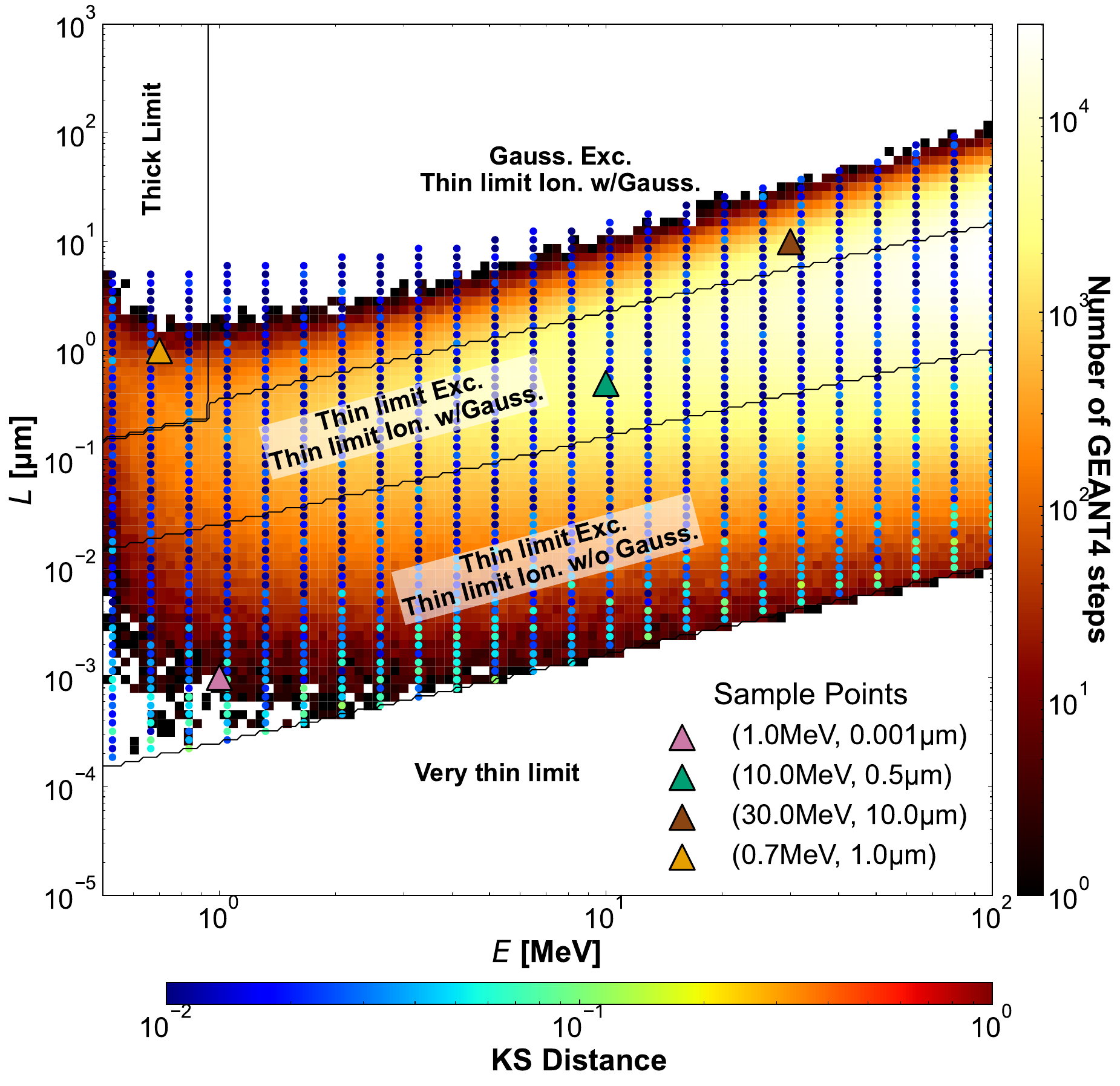}
\caption{The $E\textsf{--}L$ phase-space.
The background 2D histogram details the abundance of GEANT4 steps across the $E\textsf{--}L$ phase space, simulated with 1,000 primary protons of 100 MeV each, propagating to a full stop in aluminum. In the foreground, the grid of points used for the physics-based penalization is plotted as solid circles. The blue-to-red color-scale indicates the KS distance between the PHIN-GAN samples and the physics PDFs at the end of the training. The triangular markers indicate points, where an explicit comparison between the PHIN-GAN and the physics PDFs is presented in Fig.~\ref{fig:energy_loss_histograms}.
The same implicit division of the phase-space from Fig.~\ref{fig:pdf_phase_space} is indicated by black lines.}
\label{fig:ks_phase_space_map}
\end{figure}

\setlength{\parskip}{1em}
\setlength{\parindent}{0.5em}
We now turn to quantify the performance of the two generators in terms of the level of agreement with GEANT4, which serves as the reference standard.
In Fig.~\ref{fig:ks_phase_space_map}, we present the evaluation of the KS distances across the phase-space. 
The coordinates plotted on top of the 2D histogram of the GEANT4 steps represent the points used for physics-based penalization during training.
The marker color corresponds to the KS distance between the continuous energy-loss samples generated by the PHIN-GAN and the underlying physics PDFs at these specific points. 
The dominance of low KS values indicates a very good agreement with the underlying physics throughout the bulk of the-phase space.

\begin{figure}[!ht]
\centering    \includegraphics[width=0.45\textwidth]{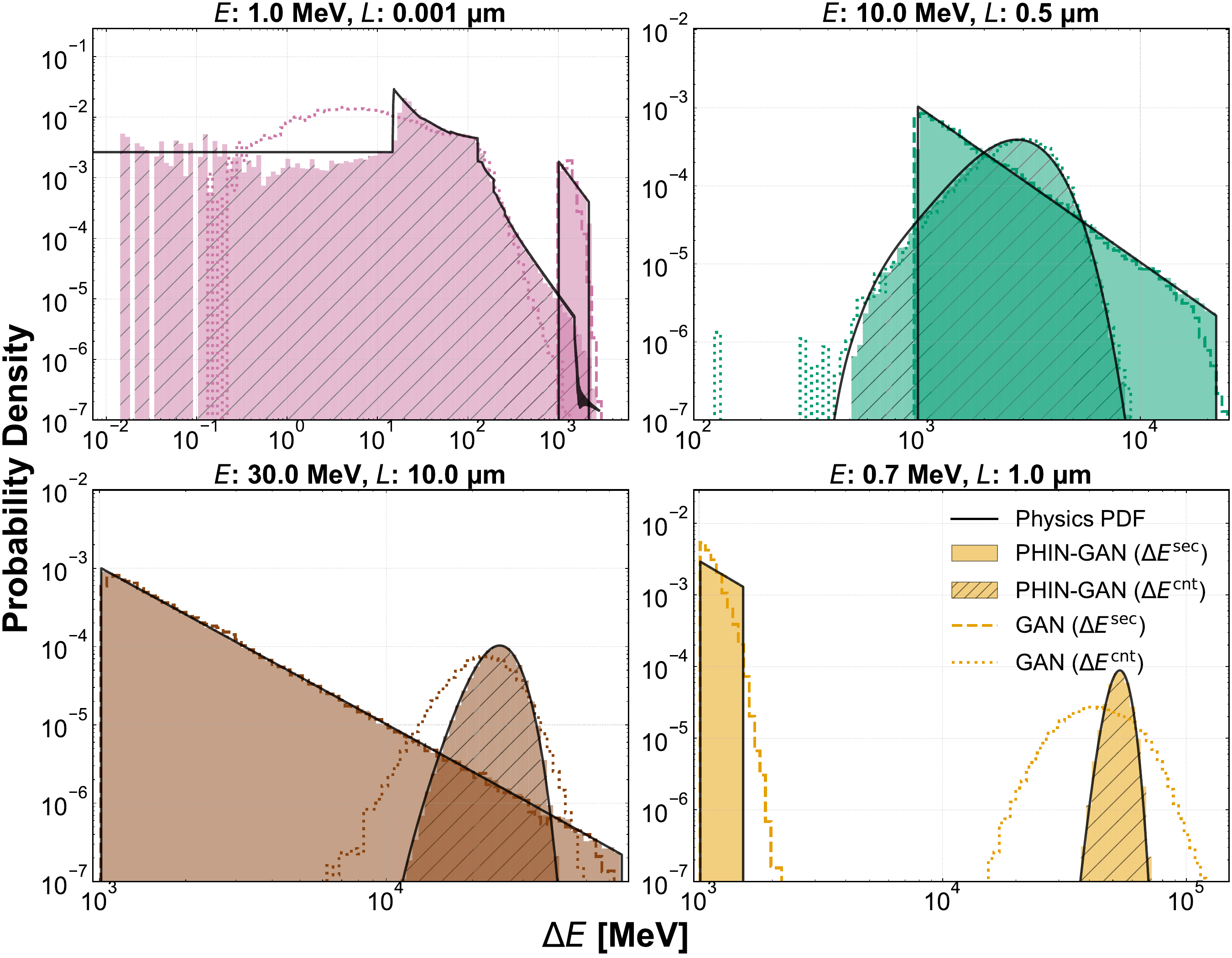}
\caption{Four example comparisons of the continuous energy-loss distributions generated by the PHIN-GAN and GAN against the physics PDFs of Sec.~\ref{sec:the-effective-straggling-function}. These distributions correspond to the four specific coordinates marked with triangles in Fig.~\ref{fig:ks_phase_space_map} (including the color code).}
\label{fig:energy_loss_histograms}
\end{figure}

To visualize the quality of the generation, Fig.~\ref{fig:energy_loss_histograms} compares a few examples of the actual energy-loss distributions produced by the PHIN-GAN and GAN against the computed physics PDFs.
These comparisons correspond to the specific coordinates marked in Fig.~\ref{fig:ks_phase_space_map}. 
It is prominent that the PHIN-GAN matches the physics PDFs and (subsequently) the GEANT4 data well, whereas the baseline GAN fails to capture the correct features.
The superior performance of PHIN-GAN in sparse regions, highlights the power of integrating physics knowledge with data-driven approaches. 
Interestingly, this integration also yields improvements in regions with abundant data.
We have checked that this good agreement shown in these four examples is kept throughout the phase-space.

\begin{figure}[!ht]
\centering    \includegraphics[width=0.45\textwidth]{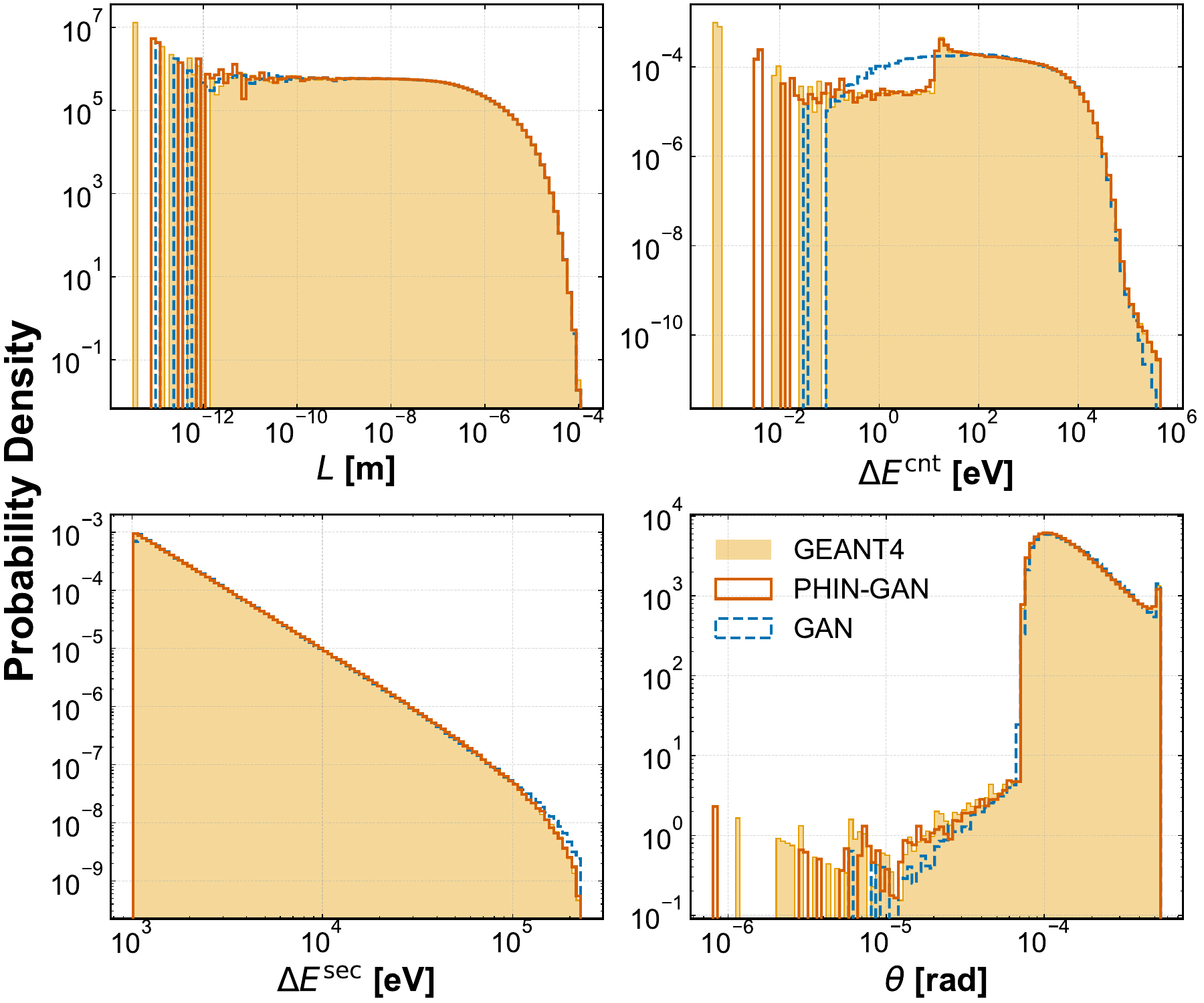}
\caption{Inclusive 1D comparison of the  samples of steps generated by GEANT4, PHIN-GAN and GAN, showing the step length (top left), continuous energy-loss (top right), secondaries energy-loss (bottom left), and scattering angle (bottom right).}
\label{fig:1d_hists}
\hfill
\end{figure}

In the next set of figures we address the level of agreement in the full, multi-dimensional phase-space.
Fig.~\ref{fig:1d_hists} presents the inclusive, 1D distributions of $L$, $\Delta E_{\rm ion}^{\rm cnt}$, $\Delta E_{\rm ion}^{\rm sec}$ and $\theta$, excluding the steps as described above.
Reflecting the shared generation algorithm for $L$, both the GAN and PHIN-GAN exhibit excellent agreement with the GEANT4 data.
For the remaining features, however, the PHIN-GAN demonstrates superior performance.
It can be clearly seen how non-trivial the distribution of $\theta$ is when MSC is ignored.
Including the MSC description simply requires the continuous module to learn one additional feature with well-defined and trivial characteristics.
While we have preliminary indications that this is indeed achievable with trivial modifications, we leave this to a future work.

\begin{figure}[!ht]
\centering    \includegraphics[width=0.45\textwidth]{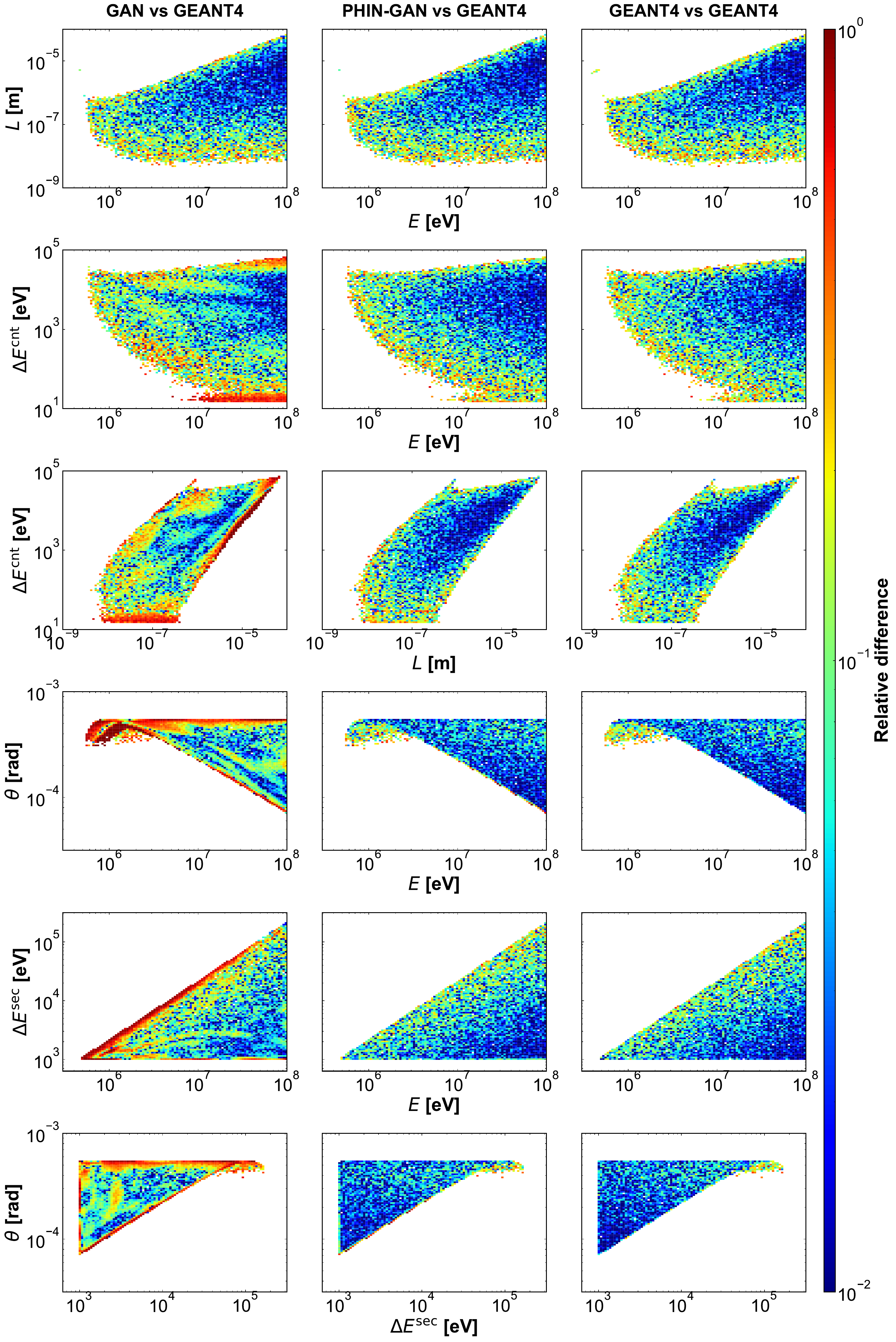}
\caption{The pair-wise relative differences for features' pairings  compared to GEANT4, including the GAN versus GEANT4 (left column), the PHIN-GAN versus GEANT4 (middle column) and GEANT4 versus GEANT4 (right column). Only slices with more than 30 steps are shown.}
\label{fig:all_relative_differences}
\hfill
\end{figure}

Crucially, while these inclusive 1D distributions provide initial validation, they are insufficient to confirm the physical realism of the generated samples.
We next provide in Fig.~\ref{fig:all_relative_differences} a qualitative examination of the features' pairwise distributions agreement.
This is illustrated via the relative differences between samples.
The ``GEANT4 versus GEANT4'' comparison (rightmost column) is included to establish a statistical baseline for the given sample size.
This reference reveals that regions, where the relative difference between PHIN-GAN and GEANT4 is noticeable are largely attributable to statistical fluctuations rather than generative misalignment.
Notably, the PHIN-GAN significantly outperforms the GAN across all regions.
It is particularly compelling that this performance gain extends to non-energy-loss related features, suggesting that the physics constraints induce a holistic improvement in the generator representation of the data.

\begin{figure}[!ht]
\centering    \includegraphics[width=0.45\textwidth]{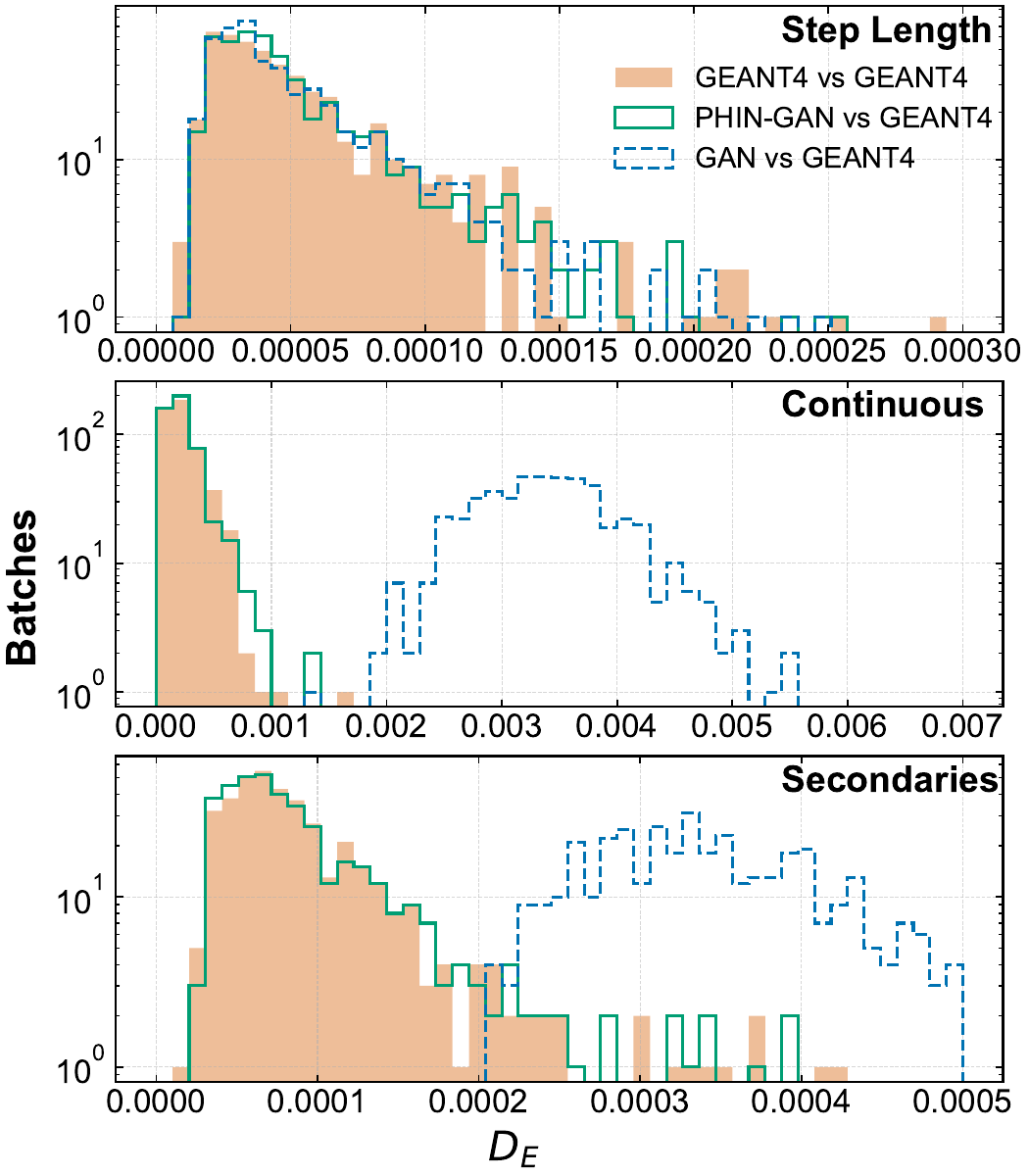}
\caption{The distributions of the ``energy distance'' metric, $D_E$, for the quantification of the multi-dimensional comparison between the three generators: GEANT4 vs GEANT4, PHIN-GAN vs GEANT4 and GAN vs GEANT4.
In the first, second and third panels the step-length, continuous and secondaries generators are evaluated, respectively.}
\label{fig:energy_distance}
\hfill
\end{figure}

To quantify the distance between the multi-dimensional distributions of the three types, we utilize the ``energy distance'' metric~\cite{rizzo2016energy} given by:
\begin{equation}
\begin{split}
D_E(X, Y) &= \frac{2}{n_x n_y} \sum_{i=1}^{n_x} \sum_{j=1}^{n_y} \|x_i - y_j\| \\
&\quad - \frac{1}{n_x^2} \sum_{i=1}^{n_x} \sum_{j=1}^{n_x} \|x_i - x_j\| \\
&\quad - \frac{1}{n_y^2} \sum_{i=1}^{n_y} \sum_{j=1}^{n_y} \|y_i - y_j\|\text{,}
\end{split}
\end{equation}
where $X$ and $Y$ are two batches of samples belonging to two simulations (e.g. GEANT4 and PHIN-GAN), $n_x$ and $n_y$ are the number of samples in each batch and $x$ and $y$ are samples of the batch. 
To avoid effects of units and features with different order of magnitudes, $D_E$ is computed over the values of the scaled features(all samples undergo the same scaling as for the PHIN-GAN discussed in Sec.~\ref{sec:physics-informed-generative-model}).
Each generator is evaluated independently as following, using four different sets.
Each set contains 490, 484 and 401 batches of 10,000 steps, for the step-length, continuous and secondaries generators respectively.
Two sets are generated by GEANT4, one by the PHIN-GAN and one by the GAN.
The steps in the two GEANT4 sets of batches are different from those the networks were trained over.
Next, the values of $D_E$ are computed between pairs of these four sets of batches (490,484 and 401).
The continuous generator batches productions are done based on the $(E, L)$ values of the steps from the first set of GEANT4 batches.
Likewise, the secondaries generator batches productions are done based on the $E'= E-\Delta E^{\rm cnt}$ values.
The $D_E$ values between the two GEANT4 sets of batches is calculated first as a reference.
Then, the $D_E$ values between the PHIN-GAN (or GAN) and GEANT4 are calculated using the second GEANT4 set of batches.
The $D_E$ histograms are shown in Fig.~\ref{fig:energy_distance}.
Using these distributions, we can perform simple KS tests and extract the corresponding $p$-values, with the ``GEANT4 vs GEANT4'' distribution serving as the reference.
    The ``PHIN-GAN vs GEANT'' $p$-value are 0.76, 0.75 and 0.94, implying that the resulting distributions are indistinguishable.
Strikingly, the $p$-values of the continuous and secondaries generators ``GAN vs GEANT4'' are practically 0, indicating a significant multi-dimensional incompatibility.

\begin{figure}[!ht]
\centering
\includegraphics[width=0.45\textwidth]{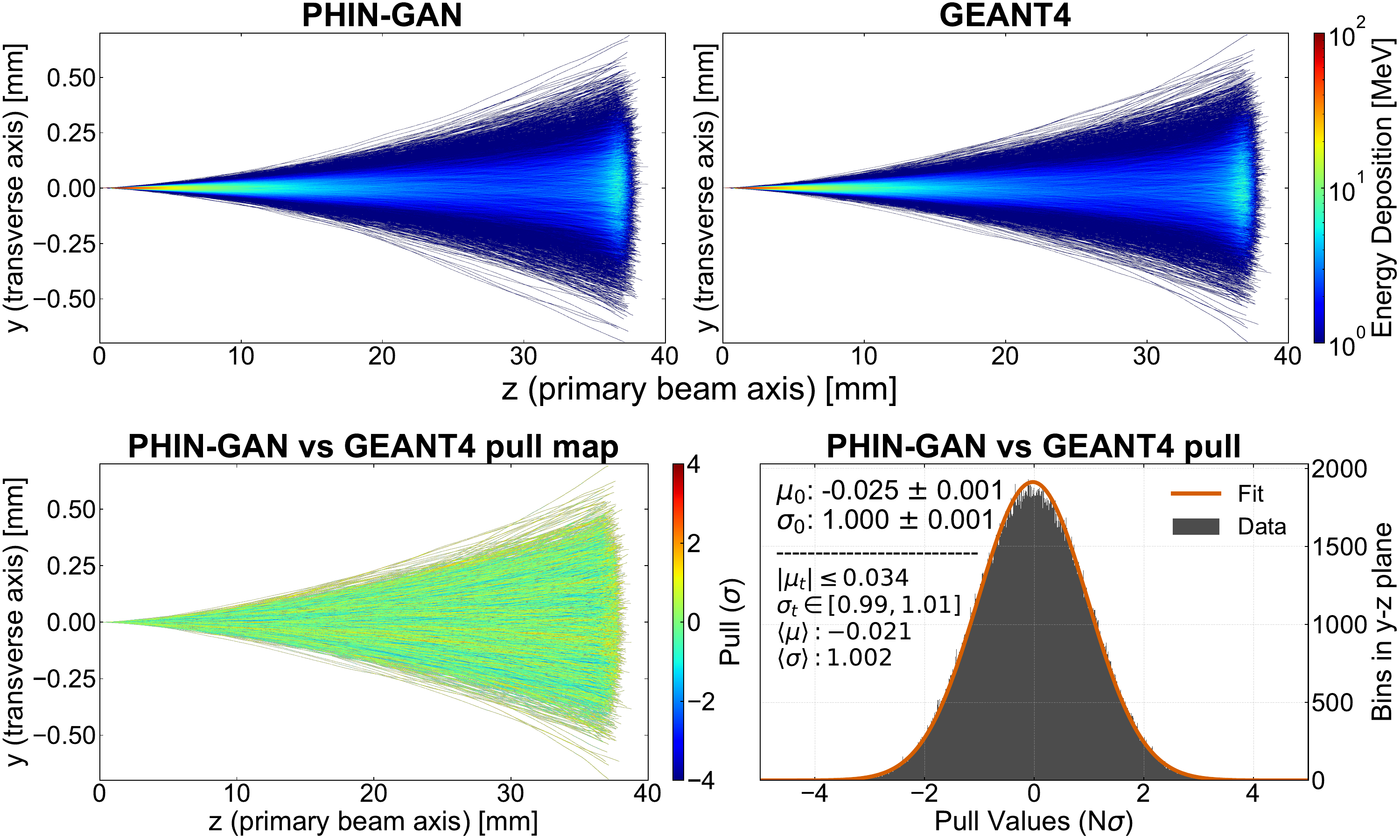}
\caption{A macroscopic comparison of the total energy-loss profiles for two samples, each comprising 10,000 proton trajectories, simulated by GEANT4 and PHIN-GAN. 
The top panels display the binned accumulated total energy-loss of one sample, each.
The bottom left panel illustrates the spatial map of the calculated pull values between the two samples (from the top panels). 
The bottom right panel presents the pull value distribution for all $y\textsf{--}z$ bins with more than three entries, overlaid with a Gaussian fit.
The fit results derived from this specific two-sample comparison are denoted as $\mu_0$ and $\sigma_0$.
The statistical errors reported by the fit are underestimated (see text) and hence the tolerances for these central values are provided below.
}
\label{fig:track_comparison}
\end{figure}

Finally, a crucial question to be asked is how small mismatches in the microscopic modeling translate to the macroscopic level behavior.
To answer this question, we look at the protons' full trajectories.
Small mismatches, if exist, are expected to accumulate throughout the full trajectory of $\sim 10^4$ steps.
Fig.~\ref{fig:track_comparison} presents a comparison of the total energy-loss ($\Delta E^{\rm tot} = \Delta E^{\rm cnt} + \Delta E^{\rm sec}$) with respect to the position of the particles simulated by GEANT4 and PHIN-GAN. 
The comparison shown here has $10,000$ primary protons for each generator.
Notably, this dataset is an order of magnitude larger than the training one.
During the PHIN-GAN simulation of full trajectories, we manually re-introduce the steps excluded from the training dataset of the continuous and secondaries modules.
The deterministic $\Delta E^{\rm cnt}$ values of steps in the respective region of the phase-space (the ``very thin limit'' discussed in Sec.~\ref{sec:the-classical-mc-sampling-approach}) are evaluated by the Bethe-Bloch formula.
The steps where $\Delta E^{\rm cnt}=0$ are re-introduced using a probability computed using our physics PDFs as explained in App.~\ref{subsec:combination-of-all-continuous-loss-processes}.
The steps with no secondary electron ($\Delta E^{\rm sec}=0$ and $\theta=0$) are re-introduced using the lookup-table based on the theoretical cross-sections as discussed in Sec.~\ref{sec:the-classical-mc-sampling-approach}.
We divide the spatial $y\textsf{--}z$ domain to a $1,500\times1,500$ grid of small bins, where $z$ is the original protons' axis and $y$ is the transverse axis. 
In each bin the total energy-loss is accumulated for all particles traversing it (top panels).
The level of agreement is quantified by computing the pull between the two simulations in each bin, defined as $\mathrm{Pull} = (a - b) / \sqrt{(\Delta a)^2 + (\Delta b)^2}$, where $a$ and $b$ represent the total energy-loss in that bin.
The uncertainties in $a$ and $b$ are calculated as $\Delta a=\sqrt{\sum_{j=1}^{N_p} (\Delta E_{j}^{\rm tot})^2}$ and likewise for $\Delta b$ in every bin, where $\Delta E_{j}^{\rm tot}$ is the total energy lost by the $j^{\rm th}$ primary proton through all steps falling in that bin and $N_p$ is the number of primary protons.
The pull values are shown in the $y\textsf{--}z$ space for a qualitative comparison and also summarized in a 1D inclusive histogram.
In the latter, 2D bins averaging fewer than 4 trajectories between the PHIN-GAN and GEANT4 samples are removed. 
This cut excludes peripheral beam regions containing isolated trajectories in only one sample; otherwise, these low-statistics bins yield artificial pull values of exactly $\pm 1$ and distort the distribution.
An unbinned maximum-likelihood fit of the pulls to a Gaussian distribution is used to check the compatibility with a normal distribution.
Both the pull calculation and the fit itself neglect the correlations between the 2D bins and hence the errors reported by the fit are underestimated.
The impact of this is negligible thanks to the fine grained binning of the 2D space.
Nevertheless, it is still possible to estimate the errors in $\mu$ and $\sigma$ by looking at independent GEANT4 trajectories' samples of similar size.
Ten such samples are generated, each of 10,000 trajectories.
By evaluating the pull distributions for all 45 unique GEANT4-GEANT4 sample pairs, an ensemble of $\{\mu_t,\sigma_t\}$ values is extracted to define the expected tolerance.
For the model evaluation, $\{\langle\mu\rangle,\langle\sigma\rangle\}$ are computed by averaging the fitted parameters of a single PHIN-GAN batch compared against each of the ten GEANT4 batches.
The generated sample is considered statistically consistent if $|\langle\mu\rangle| < \max\{|\mu_t|\}$ and $\min\{\sigma_t\}<\langle\sigma\rangle <\max\{\sigma_t\}$.
Excellent agreement is observed: the model yields $\langle\mu\rangle = -0.021$, well below the maximum GEANT4 baseline of $\max\{|\mu_t|\}=0.034$, and $\langle\sigma\rangle = 1.002$, which falls within the expected baseline interval of $\min\{\sigma_t\}=0.991$ and $\max\{\sigma_t\}=1.009$.
Despite the potential for error accumulation, the results are definitive: the PHIN-GAN captures the macroscopic distributions of GEANT4 perfectly, as indicated by the normally distributed pull values.
The macroscopic-level performance of the GAN is also tested with the same metric and following the microscopic-level miss-modeling discussed above, it is found to perform much worse.
Specifically, while $\langle\mu\rangle$ falls within the allowed range, $\langle\sigma\rangle = 1.014$ falls outside the expected range and indicates a systematically broader spread in the simulated energy-loss.
This is shown in App.~\ref{fig:gan_track_comparison}.

\begin{figure}[!ht]
\centering    \includegraphics[width=0.45\textwidth]{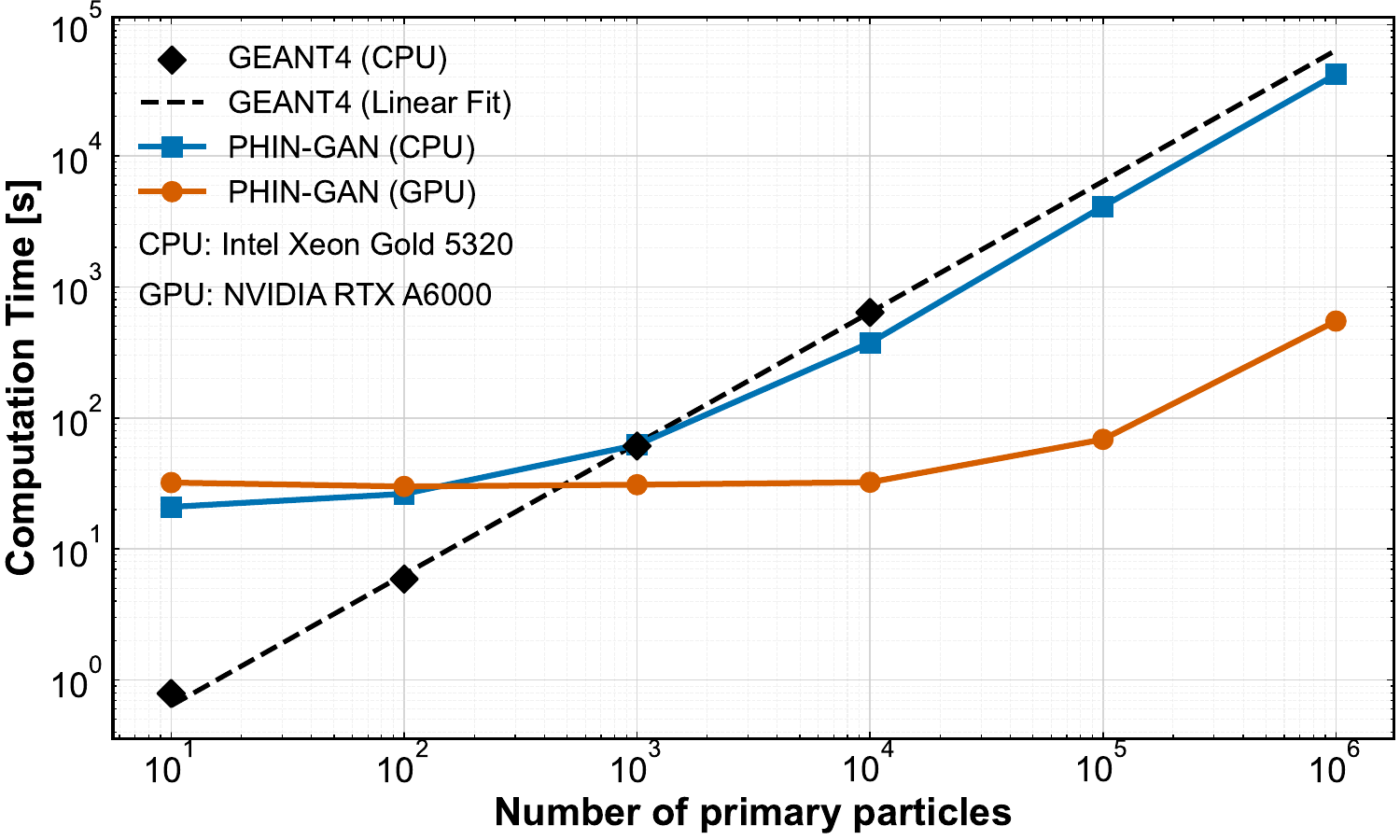}
\caption{Computation time comparison between PHIN-GAN and GEANT4 as a function of the number of primary particles simulated simultaneously. 
The black, blue and orange curves refer to simulations computed by: GEANT4 on CPU, PHIN-GAN on the same CPU and PHIN-GAN on GPU.
}
\label{fig:computation_time_simulation}
\hfill
\end{figure}

To conclude the results discussion, we at last come back to deal with the primary goal of this research: the reduction of the computation time.
Fig.~\ref{fig:computation_time_simulation} compares the computation time of the PHIN-GAN and GEANT4 for an increasing number of simulated primary particles as tested on the following devices: Intel(R) Xeon(R) Gold 5320 CPU @ 2.20GHz and NVIDIA RTX A6000 GPU.

The results highlight that running the PHIN-GAN on a GPU significantly accelerates the computation for large-scale simulations, though GEANT4 is more efficient for smaller-scale simulations. 
Overall, the PHIN-GAN achieves a massive speedup of $\times100$ relative to GEANT4, when simulating large batches (more than a 100,000 primary particles) in parallel on a GPU.
Even for smaller workloads (1,000 particles), it maintains a speedup factor of $\sim 2$.
While computation time for both GEANT4 and PHIN-GAN scales linearly with the number of steps, PHIN-GAN on a GPU exhibits highly sub-linear scaling with respect to the number of primary particles.
Hence, the speed-up factor is independent of the range-cut.
We outline that only $\sim11\%$ of the GPU usage is dedicated to the matrix multiplication of the neural networks, while the rest is occupied by the normalization layers (Pytorch's `LayerNorm' module) and the manual operations to treat the zero continuous, zero secondaries and the deterministic losses.
In this context, further optimization of the algorithm can accelerate the computation.
Lastly, we note that the particular GPU used for this benchmarking is not the most powerful available off-the-shelf, where already now there are devices faster by $\sim 3$ times.

It is important to place this GPU-based result in its current practical context.
The present typical academic computing infrastructures are mostly CPU-based.
Large-scale GPU clusters are not yet standard, and even in top-notch data centers, where many GPUs are available, the number of CPUs is substantially higher. 
However, this landscape is rapidly evolving; the trend in high-performance computing is clearly pointing towards increasing the number of GPUs to support large-scale parallel workloads.

\section{Outlook}\label{sec:outlook}
\setlength{\parskip}{1em}
\setlength{\parindent}{0.5em}

We have introduced PHIN-GAN, a physics-informed generative framework designed to reconcile computational efficiency with high-fidelity simulation.
By distilling the theoretical MC foundations of GEANT4 into continuous analytical forms, we derived the effective straggling function for a subset of dominant EM interactions across the full phase-space for the first time.
This breakthrough allows us to enforce a parametric distribution-level learning objective, ensuring the model adheres to a physics baseline.

To quantify the effectiveness of our approach and demonstrate its superiority over the same generative model
without the physics input, we subjected the two models to a rigorous benchmarking against GEANT4.
At the microscopic (step) level, a relative difference comparison verifies that the PHIN-GAN successfully reproduces the GEANT4 complex pair-wise feature correlations, while significantly outperforming the baseline GAN.
This statistical compatibility is further confirmed using the full multidimensional energy distance metric, which yields $p$-values of 0.76, 0.75 and 0.94 for the three generators, implying that the PHIN-GAN steps are practically indistinguishable from those of GEANT4.
Furthermore, this microscopic performance is translated directly to macroscopic (full trajectory) level, where the steps of one particle are stitched together to describe the full motion of the particles inside the material.
These trajectories exhibit precise agreement in the position-aware total energy-loss.
The latter is characterized by a perfect normal distributed of pulls between the PHIN-GAN and GEANT4.
Most importantly, it is noteworthy that these results are achieved with a speedup of $\times 100$, when executing large-scale batch simulations on GPUs.

While the physics PDFs are completely generic and follow the (generic) modified Urb\'an model precisely, the trained PHIN-GAN is not immediately portable, e.g., to other materials or other incoming particles.
Extending this approach to more diverse conditions requires the training of multiple models.
While this may seem prohibitive, we stress that the training has to be done only once per model and it can be envisioned that this task will be centralized by the community.

Looking ahead, grounding this POC in first principles, ensures it can be scaled and generalized to the broader collection of physical processes that are implemented in GEANT4 for the simulation of fundamental particles and radiation interaction with matter.
This paves the way for much more economic simulations at the massive scales required by both current and future experiments, without sacrificing any bit of fidelity provided by today's classical approaches.

\section{Acknowledgments}\label{sec:ack}
The work of Noam Tal Hod's group is supported by a research grant from the Estate of Dr.\ Moshe Gl\"{u}ck, the ISRAEL SCIENCE FOUNDATION (grant No. 1235/24), the Anna and Maurice Boukstein Career Development Chair, the Benoziyo Endowment Fund for the Advancement of Science, the Estate of Emile Mimran, the Estate of Betty Weneser, a research grant from the Estate of Gerald Alexander, a research grant from the Potter's Wheel Foundation, a research grant from Adam Glickman and the Sassoon \& Marjorie Peress Legacy Fund, the Deloro Center for Space and Optics.

The work of Eilam Gross's group is supported by the Minerva Grant 715027 and by Grant No 2024220 from the United States-Israel Binational Science Foundation (BSF).

Both groups are supported by the Krenter-Perinot center for High-Energy particle physics and by the Knell Family Institute for Artificial Intelligence.

\clearpage
\begingroup
\raggedright
\printbibliography

@article{ATLAS:2010arf,
    author = "Aad, G. and others",
    collaboration = "ATLAS",
    title = "{The ATLAS Simulation Infrastructure}",
    eprint = "1005.4568",
    archivePrefix = "arXiv",
    primaryClass = "physics.ins-det",
    doi = "10.1140/epjc/s10052-010-1429-9",
    journal = "Eur. Phys. J. C",
    volume = "70",
    pages = "823--874",
    year = "2010"
}

@article{Hashemi_2024,
   title={Deep generative models for detector signature simulation: A taxonomic review},
   volume={12},
   ISSN={2405-4283},
   url={http://dx.doi.org/10.1016/j.revip.2024.100092},
   DOI={10.1016/j.revip.2024.100092},
   journal={Reviews in Physics},
   publisher={Elsevier BV},
   author={Hashemi, Baran and Krause, Claudius},
   year={2024},
   month=dec, pages={100092} }

@Inbook{Iacus2018,
author="Iacus, Stefano M.
and Yoshida, Nakahiro",
title="Compound Poisson Processes",
bookTitle="Simulation and Inference for Stochastic Processes with YUIMA: A Comprehensive R Framework for SDEs and Other Stochastic Processes",
year="2018",
publisher="Springer International Publishing",
address="Cham",
pages="137--154",
isbn="978-3-319-55569-0",
doi="10.1007/978-3-319-55569-0_3",
url="https://doi.org/10.1007/978-3-319-55569-0_3"
}

@techreport{CERN-LHCC-2022-005,
      collaboration = "ATLAS",
      title         = "{ATLAS Software and Computing HL-LHC Roadmap}",
      institution   = "CERN",
      reportNumber  = "CERN-LHCC-2022-005, LHCC-G-182",
      address       = "Geneva",
      year          = "2022",
      url           = "https://cds.cern.ch/record/2802918",
}

@techreport{CERN-LHCC-2020-015,
      collaboration = "ATLAS",
      title         = "{ATLAS HL-LHC Computing Conceptual Design Report}",
      institution   = "CERN",
      reportNumber  = "CERN-LHCC-2020-015, LHCC-G-178",
      address       = "Geneva",
      year          = "2020",
      url           = "https://cds.cern.ch/record/2729668",
}

@article{Campana:2018,
    author = "Campana, Simone and Bird, Ian and Panzer-Steindel, Bernd",
    title = "{Overview of the WLCG strategy towards HL-LHC}",
    doi = "10.1051/epjconf/201921402001",
    journal = "EPJ Web Conf.",
    volume = "214",
    pages = "02001",
    year = "2019"
}

@misc{CERN_WLCG,
  author       = {{CERN}},
  title        = {{The Worldwide LHC Computing Grid (WLCG)}},
  howpublished = {\url{https://home.web.cern.ch/science/computing/grid}},
  year         = {2026},
  note         = {Accessed: 2026-01-08}
}

@article{Landau:216256,
      author        = "Landau, Lev Davidovich",
      title         = "{On the energy loss of fast particles by ionization}",
      journal       = "J. Phys.",
      volume        = "8",
      number        = "4",
      pages         = "201-205",
      year          = "1944",
      url           = "https://cds.cern.ch/record/216256",
}

@article{bichsel1988straggling,
  title={Straggling in thin silicon detectors},
  author={Bichsel, Hans},
  journal={Reviews of Modern Physics},
  volume={60},
  number={3},
  pages={663},
  year={1988},
  publisher={APS}
}

@article{osti_4311507,
  author       = {Vavilov, P V},
  title        = {IONIZATION LOSSES OF HIGH-ENERGY HEAVY PARTICLES},
  url          = {https://www.osti.gov/biblio/4311507},
  journal      = {Soviet Phys. JETP},
  volume       = {Vol: 5},
  place        = {Country unknown/Code not available},
  year         = {1957},
  month        = {11}}

@article{agostinelli2003geant4,
  title={GEANT4—a simulation toolkit},
  author={Agostinelli, Sea and Allison, John and Amako, K al and Apostolakis, John and Araujo, H and Arce, Pedro and Asai, Makoto and Axen, D and Banerjee, Swagato and Barrand, GJNI and others},
  journal={Nuclear instruments and methods in physics research section A: Accelerators, Spectrometers, Detectors and Associated Equipment},
  volume={506},
  number={3},
  pages={250--303},
  year={2003},
  publisher={Elsevier}
}

@article{1610988,
  author={Allison, J. and Amako, K. and Apostolakis, J. and Araujo, H. and Arce Dubois, P. and Asai, M. and Barrand, G. and Capra, R. and Chauvie, S. and Chytracek, R. and Cirrone, G.A.P. and Cooperman, G. and Cosmo, G. and Cuttone, G. and Daquino, G.G. and Donszelmann, M. and Dressel, M. and Folger, G. and Foppiano, F. and Generowicz, J. and Grichine, V. and Guatelli, S. and Gumplinger, P. and Heikkinen, A. and Hrivnacova, I. and Howard, A. and Incerti, S. and Ivanchenko, V. and Johnson, T. and Jones, F. and Koi, T. and Kokoulin, R. and Kossov, M. and Kurashige, H. and Lara, V. and Larsson, S. and Lei, F. and Link, O. and Longo, F. and Maire, M. and Mantero, A. and Mascialino, B. and McLaren, I. and Mendez Lorenzo, P. and Minamimoto, K. and Murakami, K. and Nieminen, P. and Pandola, L. and Parlati, S. and Peralta, L. and Perl, J. and Pfeiffer, A. and Pia, M.G. and Ribon, A. and Rodrigues, P. and Russo, G. and Sadilov, S. and Santin, G. and Sasaki, T. and Smith, D. and Starkov, N. and Tanaka, S. and Tcherniaev, E. and Tome, B. and Trindade, A. and Truscott, P. and Urban, L. and Verderi, M. and Walkden, A. and Wellisch, J.P. and Williams, D.C. and Wright, D. and Yoshida, H.},
  journal={IEEE Transactions on Nuclear Science},
  title={Geant4 developments and applications},
  year={2006},
  volume={53},
  number={1},
  pages={270-278},
  doi={10.1109/TNS.2006.869826}
}

@article{ALLISON2016186,
	author = {J. Allison and K. Amako and J. Apostolakis and P. Arce and M. Asai and T. Aso and E. Bagli and A. Bagulya and S. Banerjee and G. Barrand and B.R. Beck and A.G. Bogdanov and D. Brandt and J.M.C. Brown and H. Burkhardt and Ph. Canal and D. Cano-Ott and S. Chauvie and K. Cho and G.A.P. Cirrone and G. Cooperman and M.A. Cort{\'e}s-Giraldo and G. Cosmo and G. Cuttone and G. Depaola and L. Desorgher and X. Dong and A. Dotti and V.D. Elvira and G. Folger and Z. Francis and A. Galoyan and L. Garnier and M. Gayer and K.L. Genser and V.M. Grichine and S. Guatelli and P. Gu{\`e}ye and P. Gumplinger and A.S. Howard and I. H{\v r}ivn{\'a}{\v c}ov{\'a} and S. Hwang and S. Incerti and A. Ivanchenko and V.N. Ivanchenko and F.W. Jones and S.Y. Jun and P. Kaitaniemi and N. Karakatsanis and M. Karamitros and M. Kelsey and A. Kimura and T. Koi and H. Kurashige and A. Lechner and S.B. Lee and F. Longo and M. Maire and D. Mancusi and A. Mantero and E. Mendoza and B. Morgan and K. Murakami and T. Nikitina and L. Pandola and P. Paprocki and J. Perl and I. Petrovi{\'c} and M.G. Pia and W. Pokorski and J.M. Quesada and M. Raine and M.A. Reis and A. Ribon and A. {Risti{\'c} Fira} and F. Romano and G. Russo and G. Santin and T. Sasaki and D. Sawkey and J.I. Shin and I.I. Strakovsky and A. Taborda and S. Tanaka and B. Tom{\'e} and T. Toshito and H.N. Tran and P.R. Truscott and L. Urban and V. Uzhinsky and J.M. Verbeke and M. Verderi and B.L. Wendt and H. Wenzel and D.H. Wright and D.M. Wright and T. Yamashita and J. Yarba and H. Yoshida},
	journal = {Nuclear Instruments and Methods in Physics Research Section A: Accelerators, Spectrometers, Detectors and Associated Equipment},
	pages = {186-225},
	title = {Recent developments in Geant4},
	volume = {835},
	year = {2016}
}

@article{G4MAN,
    Author = {GEANT4 Physics Reference Manual, v11.2},
    journal = {\url{https://geant4-userdoc.web.cern.ch/UsersGuides/PhysicsReferenceManual/fo/PhysicsReferenceManual.pdf}},
    url = {https://geant4-userdoc.web.cern.ch/UsersGuides/PhysicsReferenceManual/fo/PhysicsReferenceManual.pdf},
    doi = {},
    year = {}
}

@inproceedings{arjovsky2017wasserstein,
  title={Wasserstein generative adversarial networks},
  author={Arjovsky, Martin and Chintala, Soumith and Bottou, L{\'e}on},
  booktitle={International conference on machine learning},
  pages={214--223},
  year={2017},
  organization={PMLR}
}

@article{gulrajani2017improved,
  title={Improved training of wasserstein gans},
  author={Gulrajani, Ishaan and Ahmed, Faruk and Arjovsky, Martin and Dumoulin, Vincent and Courville, Aaron C},
  journal={Advances in neural information processing systems},
  volume={30},
  year={2017}
}

@article{mirza2014conditional,
  title={Conditional generative adversarial nets},
  author={Mirza, Mehdi},
  journal={arXiv preprint arXiv:1411.1784},
  year={2014}
}

@article{lassila1995energy,
  title={Energy loss in thin layers in GEANT},
  author={Lassila-Perini, Kati and Urb{\'a}n, L{\'a}szl{\'o}},
  journal={Nuclear Instruments and Methods in Physics Research Section A: Accelerators, Spectrometers, Detectors and Associated Equipment},
  volume={362},
  number={2-3},
  pages={416--422},
  year={1995},
  publisher={Elsevier}
}

@article{rizzo2016energy,
  title={Energy distance},
  author={Rizzo, Maria L and Sz{\'e}kely, G{\'a}bor J},
  journal={wiley interdisciplinary reviews: Computational statistics},
  volume={8},
  number={1},
  pages={27--38},
  year={2016},
  publisher={Wiley Online Library}
}

@inproceedings{johnson2024celeritas,
  title={Celeritas: Accelerating geant4 with gpus},
  author={Johnson, Seth R and Esseiva, Julien and Biondo, Elliott and Canal, Philippe and Demarteau, Marcel and Evans, Thomas and Jun, Soon Yung and Lima, Guilherme and Lund, Amanda and Romano, Paul and others},
  booktitle={EPJ Web of Conferences},
  volume={295},
  pages={11005},
  year={2024},
  organization={EDP Sciences}
}

@article{ParticleDataGroup:2024cfk,
    author = "Navas, S. and others",
    collaboration = "Particle Data Group",
    title = "{Review of particle physics}",
    doi = "10.1103/PhysRevD.110.030001",
    journal = "Phys. Rev. D",
    volume = "110",
    number = "3",
    pages = "030001",
    year = "2024"
}

@misc{1370017282431050757,
author="Tieleman, T.",
title="Lecture 6.5‐rmsprop: Divide the Gradient by a Running Average of Its Recent Magnitude",
journal="COURSERA: Neural Networks for Machine Learning",
year="2012",
volume="4",
number="2",
pages="26",
URL="https://cir.nii.ac.jp/crid/1370017282431050757"
}

@misc{g4fastsim,
  author       = {{Geant4 Collaboration}},
  title        = {{Geant4 Fast Simulation}},
  howpublished = {\url{https://g4fastsim.web.cern.ch/}},
}

@article{perl2012topas,
  title={TOPAS: an innovative proton Monte Carlo platform for research and clinical applications},
  author={Perl, J and Shin, J and Schumann, J and Faddegon, B and Paganetti, H},
  journal={Medical physics},
  volume={39},
  number={11},
  pages={6818--6837},
  year={2012},
  publisher={American Association of Physicists in Medicine}
}

@article{jan2004gate,
  title={GATE: a simulation toolkit for PET and SPECT},
  author={Jan, Sebastien and Santin, Giovanni and Strul, Daniel and Staelens, Steven and Assi{\'e}, K and Morel, Christian and others},
  journal={Physics in Medicine \& Biology},
  volume={49},
  number={19},
  pages={4543},
  year={2004},
  publisher={IOP Publishing}
}

@article{incerti2010geant4,
  title={The Geant4-DNA project},
  author={Incerti, S and Baldacchino, G and Bernal, M and Capra, R and Champion, C and Francis, Z and others},
  journal={International Journal of Modeling, Simulation, and Scientific Computing},
  volume={1},
  number={02},
  pages={157--178},
  year={2010},
  publisher={World Scientific}
}

@article{morishima2017discovery,
  title={Discovery of a big void in Khufu’s Pyramid by observation of cosmic-ray muons},
  author={Morishima, Kunihiro and Kuno, Mitsuaki and Nishio, Akira and Kitagawa, Nobuko and Manabe, Yuta and Moto, Masaki and Takasaki, Fumihiko and Fujii, Hirofumi and Satoh, Kotaro and Kodama, Hideyo and others},
  journal={Nature},
  volume={552},
  number={7685},
  pages={386--390},
  year={2017},
  publisher={Nature Publishing Group UK London}
}

@article{hohlmann2009geant4,
  title={GEANT4 simulation of a cosmic ray muon tomography system with micro-pattern gas detectors for the detection of high-Z materials},
  author={Hohlmann, Marcus and Ford, Patrick and Gnanvo, Kondo and Helsby, Jennifer and Pena, David and Hoch, Richard and Mitra, Debasis},
  journal={IEEE Transactions on Nuclear Science},
  volume={56},
  number={3},
  pages={1356--1363},
  year={2009},
  publisher={IEEE}
}

@article{atwood2009large,
  title={The large area telescope on the Fermi gamma-ray space telescope mission},
  author={Atwood, William B and Abdo, Aous A and Ackermann, Markus and Althouse, W and Anderson, B and Axelsson, M and Baldini, Luca and Ballet, J and Band, DL and Barbiellini, Guido and others},
  journal={The Astrophysical Journal},
  volume={697},
  number={2},
  pages={1071},
  year={2009},
  publisher={IOP Publishing}
}

@article{aad2008atlas,
  title={The ATLAS experiment at the CERN large hadron collider},
  author={Aad, Georges and Anduaga, Xabier Sebastian and Antonelli, S and Bendel, M and Breiler, B and Castrovillari, F and Civera, JV and Del Prete, T and Dova, Maria Teresa and Duffin, S and others},
  year={2008},
  publisher={IOP Publishing}
}

@article{cms2008cms,
  title={The CMS experiment at the CERN LHC},
  author={Cms Collaboration and others},
  journal={Journal of instrumentation},
  volume={3},
  number={August 2008},
  pages={1--334},
  year={2008},
  publisher={IOP Publishing}
}

@article{aubert2002babar,
  title={The BABAR detector},
  author={Aubert, Bernard and Bazan, A and Boucham, A and Boutigny, D and De Bonis, I and Favier, J and Gaillard, J-M and Jeremie, A and Karyotakis, Y and Le Flour, T and others},
  journal={Nuclear Instruments and Methods in Physics Research Section A: Accelerators, Spectrometers, Detectors and Associated Equipment},
  volume={479},
  number={1},
  pages={1--116},
  year={2002},
  publisher={Elsevier}
}

@article{belle2002belle,
  title={The belle detector},
  author={Belle Collaboration and others},
  journal={Nuclear Instruments and Methods in Physics Research, Section A: Accelerators, Spectrometers, Detectors and Associated Equipment},
  volume={479},
  number={1},
  pages={117--232},
  year={2002},
  publisher={Elsevier BV}
}

@article{schael2006precision,
  title={Precision electroweak measurements on the Z resonance},
  author={Schael, Stefan and Barate, R and Bruneliere, R and Buskulic, D and De, BI and Decamp, D and Ghez, P and Goy, C and Jezequel, S and Lees, JP and others},
  year={2006},
  publisher={Elsevier}
}

@techreport{apollinari2016high,
  title={High-luminosity large Hadron collider (HL-LHC). Technical design report V. 0.1},
  author={Apollinari, Giorgio and B{\'e}jar Alonso, I and Br{\"u}ning, O and Fessia, P and Lamont, M and Rossi, L and Tavian, L},
  year={2016},
  institution={Fermi National Accelerator Laboratory (FNAL), Batavia, IL (United States)}
}

@article{abada2019fcc,
  title={FCC physics opportunities},
  author={Abada, Asmaa and Abbrescia, Marcello and AbdusSalam, Shehu S and Abdyukhanov, I and Fernandez, J Abelleira and Abramov, A and Aburaia, Mohamed and Acar, AO and Adzic, PR and Agrawal, Prateek and others},
  journal={The European Physical Journal C},
  volume={79},
  number={6},
  pages={1--161},
  year={2019},
  publisher={Springer}
}

@article{khalek2022science,
  title={Science requirements and detector concepts for the electron-ion collider: EIC yellow report},
  author={Khalek, R Abdul and Accardi, A and Adam, J and Adamiak, D and Akers, W and Albaladejo, M and Al-Bataineh, A and Alexeev, MG and Ameli, F and Antonioli, P and others},
  journal={Nuclear Physics A},
  volume={1026},
  pages={122447},
  year={2022},
  publisher={Elsevier}
}

@article{black2024muon,
  title={Muon collider forum report},
  author={Black, KM and Jindariani, S and Li, D and Maltoni, F and Meade, P and Stratakis, D and Acosta, D and Agarwal, R and Agashe, K and Aim{\`e}, C and others},
  journal={Journal of Instrumentation},
  volume={19},
  number={02},
  pages={T02015},
  year={2024},
  publisher={IOP Publishing}
}
\endgroup
\newpage

\clearpage
\onecolumn 
\appendix  

\numberwithin{equation}{section}
\numberwithin{figure}{section}

\section{The modified-Urb\'an model}\label{sec:appendix_a}
\setlength{\parskip}{1em}
\setlength{\parindent}{0.5em}

\subsection{Overview}
\label{subsec:mathematical_modeling_overview}

Charged particles propagating through matter lose energy through a combination of EM mechanisms, where excitation and ionization are the most abundant ones.
GEANT4 implements these mechanisms through MC simulations.
To optimize computational performance, ionization events are separated into low- and high-energy transfer regimes defined by the cutoff $T_{\rm cut}$.
The first, along with excitations, are categorized as continuous energy-loss while the latter discretely ends a step with the generation of a $\delta$-electron.
For simplification, we will refer the latter as secondary energy-loss hereafter.

In this section, we introduce and document the highly successful modified-Urb\'an model as implemented in
GEANT4, which serves as the foundation for its EM energy-loss fluctuation simulations.
This model has been instrumental in accurately representing the straggling of charged particles in a wide range of
materials and experimental setups.
The following is primarily based on the GEANT4 physics reference manual~\cite{G4MAN} and closely follows the
Urb\'an model~\cite{lassila1995energy}.
To date, the specific implementation of \texttt{G4UniversalFluctuation} remains undocumented.

\subsection{Fluctuations in Thick Absorbers}
\label{subsec:thick_absorber}

The total continuous energy-loss of charged particles is inherently stochastic, characterized by a distribution known as the straggling function. 
In the ``thick absorber'' regime, the particle undergoes a sufficiently large number of collisions such that the statistical fluctuations are dominated by the Central Limit Theorem (CLT).
Additionally, the atomic structure (binding energies) is neglected, allowing the scattering process to be modeled as interactions with free electrons.
Consequently, the straggling function in this regime converges to a Gaussian distribution.

In GEANT4, this Gaussian regime is identified using a lower limit on the number of interactions, parameterized by $\kappa$ (default $\kappa = 10$).
For thick absorbers where a high range-cut is applied (implying energy-loss without explicit $\delta$-ray production), the condition
\begin{equation}
\label{eq:app_high_range_cut}
\left<\Delta E\right> > \kappa T_{\max}
\end{equation}
must hold.
Here, $T_{\max}$ is the maximum kinetic energy transferable to an atomic electron:
\begin{equation}
\label{eq:app_T_max}
T_{\max} = \frac{2m_e\beta^2\gamma^2}{1+2\gamma m_e/m + (m_e/m)^2}\text{, }
\end{equation}
where $m_e$ is the electron mass, $m$ is the primary particle mass, and $\beta, \gamma$ are the standard relativistic factors.

In most simulations, secondary $\delta$-electrons above a threshold $T_{\rm cut}$ are generated explicitly. In this case, the continuous energy-loss is "restricted" to transfers below $T_{\rm cut}$. The condition for the Gaussian approximation from Eq.~\eqref{eq:app_high_range_cut} is updated to $\left<\Delta E\right> > \kappa T_{\rm cut}$ provided that the maximum transfer is not excessively larger than the cut:
\begin{equation}
\label{eq:app_max_transfer_limit}
T_{\max} \leq 2 T_{\rm cut}.
\end{equation}

Under these restricted conditions, the straggling function approaches a Gaussian distribution, with mean energy-loss, $\langle \Delta E \rangle$, determined by the Bethe-Bloch formula\footnote{This
formulation of the Bethe-Bloch is valid for protons, $\alpha$ particles and atomic ions.
For electrons, further corrections to the equation is necessary due to their small mass, indistinguishabillity and
Bremsstrahlung interactions~\cite{ParticleDataGroup:2024cfk}.} and variance given by Bohr's formula for free electrons:
\begin{equation}
\label{eq:app_bethe_bloch}
    -\left< \frac{dE}{dx} \right> = \frac{4 \pi}{m_e c^{2}} \cdot \frac{N_{e} z^{2}}{\beta^{2}} \cdot
\left( \frac{e^{2}}{4 \pi \epsilon_{0}} \right)^{2} \cdot \left[ \ln \left( \frac{2 m_{e}
c^{2} \beta^{2}}{I \cdot \left( 1-\beta^{2} \right) } \right) -\beta^{2} \right]\text{, }
\end{equation}
\begin{equation}
\Omega^2 = 2\pi r_e^2 m_e c^2 N_{e} Z_h^2 \beta^2 T_{\max} s \left(1 - \frac{\beta^2}{2} \frac{T_{\rm cut}}{T_{\max}}\right)\label{eq:app_bohr_variance}\text{, }
\end{equation}
where $N_{e}$ is the electron density of the medium, \( z \) is the charge of the incident particle, \( c \) is the speed of light, \( e \) is the elementary charge, $I$ is the mean excitation energy and \( \epsilon_0 \) is the vacuum permittivity, $r_e$ is the classical electron radius, $s$ is the step length and $Z_h$ is the charge of the incident particle in units of positron charge.

In case $\mathcal{E} \equiv \frac{\langle \Delta E_{\rm ion} \rangle}{\Omega} < 2$, where $\langle \Delta E_{\rm ion} \rangle$ is the mean energy
loss due to ionization, loss is sampled from $\Gamma \left( \mathcal{E}^2 \right)$ instead of a Gaussian.
In our simulation environment this condition is never met~\footnote{This limit is irrelevant to the phase-space in the discussion below, but for completeness it will be implemented at a later time.}.

\subsection{Fluctuations in Thin Absorbers}
\label{subsec:thin_absorber}
The primary motivation for employing the modified-Urb\'an model over standard transport theories (such as Landau or Vavilov-Landau distributions) lies in the breakdown of the 'free electron' approximation within thin absorbers. 
In this regime, the energy-loss is often comparable to the atomic binding energies of the material, breaking the free electrons approximation.
The Urb\'an model addresses this by phenomenologically decomposing the energy-loss into two distinct physical channels: excitation and ionization.
Excitation accounts for the quantized, resonant nature of interactions with bound atomic shells (approximated here as fixed energy levels), while ionization captures the stochastic, continuous spectrum of higher energy transfers governed by the $1/E^2$ Rutherford cross-section.

Beyond the physical decomposition, the model employs a statistical strategy to handle the heavy-tailed nature of ionization energy-loss. 
In any given step, a particle undergoes a vast number of 'soft' (low-energy) collisions and a small, stochastic number of 'hard' (high-energy) collisions.
Simulating every soft collision is computationally prohibitive and unnecessary, as their aggregate contribution converges to a Gaussian distribution via the Central Limit Theorem.
Conversely, the rare hard collisions deviate from this convergence and are entirely responsible for the asymmetric 'Landau tail' of the straggling function. To balance accuracy with efficiency, the modified-Urb\'an model introduces a dynamic energy threshold.
Collisions below this threshold are aggregated into a Gaussian term, while the few collisions above it are explicitly resolved as individual collisions within a compound Poisson process.

To physically realize this decomposition, the approach uses a simplified atomic model, assuming atoms possess only two discrete energy levels, each defined by binding energies $E_1$ and $E_2$.
When a charged particle interacts with the atom, energy-loss may occur via excitation (with associated energy-losses $E_1$ or $E_2$) or through ionization, where energy-loss is governed by a density function $g(E) \sim \frac{1}{E^2}$ realized from the Rutherford cross-section.
$g(E)$ is normalized as follows:
\begin{equation}
    \int_{A}^{B} g(E) \, dE = 1 \Rightarrow g(E) = \frac{A B}{B - A} \frac{1}{E^2} \text{, }\label{eq:app_density_function}
\end{equation}
where for continuous ionization processes $A = \alpha E_0=w_3$ and $B = T_{\rm up} = \min(T_{\max}, T_{\rm cut})$, while for generation of secondaries ($\delta$ electrons) $A = T_{\text{cut}}$ and $B = T_{\max}$.
$\alpha$ is the dynamic variable, and accordingly, $w_3$ is the dynamic energy threshold.
Their mathematical expressions are described in~\ref{sec:appendix_a_ionization}.

The model’s macroscopic cross-sections for excitation, \( \Sigma_i \) (for \( i = 1, 2 \)), are expressed as:

\begin{equation}
\label{eq:app_excitation_cross_section}
    \Sigma_i = C \frac{f_i}{E_i} \ln \left( \frac{2mc^2 (\beta \gamma)^2}{E_i} \right) - \beta^2 \ln \left( \frac{2mc^2 (\beta \gamma)^2}{I} \right) - \beta^2 (1 - r),
\end{equation}

whereas the ionization cross-section \( \Sigma_3 \) is given by:

\begin{equation}
    \Sigma_3 = C \frac{T_{\text{up}} - E_0}{E_0 T_{\text{up}}} \ln \left( \frac{T_{\text{up}}}{E_0} \right) r.
\end{equation}
Here, \( I \) is the mean ionization energy, \( r \) is the fraction of ionization component with respect to excitation
and \( E_i \) and \( f_i \) denote the energy levels and their corresponding oscillator strengths.
\( C \) is a parameter of the model, with \( f_i \) and \( E_i \) constrained by:

\begin{equation}
\label{eq:app_osc_and_energy_constrains}
    f_1 + f_2 = 1, \quad f_1 \cdot \ln E_1 + f_2 \cdot \ln E_2 = \ln I.
\end{equation}

The parameter \( C \) is derived from the mean number of collisions \( \langle n_i \rangle \) over a distance
\( L \) as \( \langle n_i \rangle = L \Sigma_i \), where the mean energy-loss of continuous ionization in a
given step \( L \) is:

\begin{equation}
    \left< \frac{dE}{dx} \right> \cdot L = \left( \Sigma_1 E_1 + \Sigma_2 E_2 + \int_{E_0}^{T_{\text{up}}} E g(E) \, dE \right) L.
\end{equation}
From equations ~\eqref{eq:app_excitation_cross_section} through~\eqref{eq:app_osc_and_energy_constrains}, it follows that \( C = \langle dE/dx \rangle \).

In setting the fluctuation model parameters, the values are assigned to match specific atomic characteristics.
The quantities \( Z \cdot f_1 \) and \( Z \cdot f_2 \) correspond to loosely and tightly bound electrons, respectively.
Specifically, \( f_2 = 0 \) for \( Z = 1 \) and \( f_2 = \frac{2}{Z} \) for \( Z \geq 2 \), while \( E_2 \approx 10 \, \text{eV} \times Z^2 \)
corresponds to the K-shell energy level.
For ionization energy, \( E_0 \approx 10 \, \text{eV} \).
Using these definitions, the parameters \( f_1 \) and \( E_1 \) are derived from equations~\eqref{eq:app_osc_and_energy_constrains}.
The model parameter \( r \), which dictates the relative contributions of ionization and excitation to total energy-loss,
has been calibrated to \( r = 0.56 \) based on empirical comparisons between simulated and experimental energy-loss distributions.
The current implementation in GEANT4 considers just one excitation level, by setting \( f_2 \) and \( E_2 \) to 0.

This model forms a foundational component of energy straggling studies, particularly relevant for high-precision
simulations of energy-loss distributions in thin materials, which is essential in understanding radiation interactions
at microscopic levels.
It is important to emphasize that this model serves as a basis for generating MC simulations but does not offer an
analytical probabilistic description of the energy-loss processes.

\subsection{Monte-Carlo sampling of energy-loss}\label{subsec:monte-carlo-sampling-of-energy-loss}

The MC sampling algorithm of energy-loss in GEANT4 is a sequential algorithm that accounts for the various energy-loss
mechanisms.

Initially, it is first checked if the mean energy-loss ($\Delta E = \left< \frac{dE}{dx} \right> \cdot L$) is below a
fixed threshold of $10$~eV\@.
If this condition is met, the energy-loss is simply fixed to the mean loss, given by the Bethe-Bloch~\eqref{eq:app_bethe_bloch}.

Next, the `thick' absorber limit conditions are checked (equations \ref{eq:app_high_range_cut} and~\ref{eq:app_max_transfer_limit}).
If these conditions are met, the energy-loss is sampled from a Gaussian distribution.
Otherwise, the `thin' absorber limit is applied, where the energy-loss is computed explicitly by combining excitation and ionization processes, sampled as follows:
\begin{itemize}
    \item \textbf{Excitation:} A complex mixture of Poisson and uniform distributions for low excitation collision rate, or a truncated Gaussian distribution for high.
    \item \textbf{Ionization:} A `compound Poisson process'~\cite{Iacus2018}, with an additional truncated Gaussian fluctuation for large mean numbers of collisions.
\end{itemize}

Finally, when an ionization collision occurs, where the energy transferred to the target electron is above $T_{\rm cut}$, the target electron is explicitly tracked as a secondary electron and an additional fluctuation term is added to the energy-loss of the primary particle.

In the following subsection we will describe the details of the sampling of energy-loss based on the continuous excitation,
continuous ionization and secondary production, as computed by GEANT4.

\subsubsection{Excitations} \label{subsubsec:excitations}

From computational efficiency reasons, the model distinguishes between high and low collision rate excitation regimes for the computation of the total energy-loss due to excitations, denoted $\Delta E_{\rm exc}$.

When the mean number of excitation collisions, $n_{\rm exc}^1$ and $n_{\rm exc}^2$, exceeds $n_{\max}^{\rm exc}$ (threshold, fixed to 8), the energy-loss is sampled from a Gaussian distribution with mean and variance:
\begin{equation}
    \label{eq:app_gaussian_excitation_mean}
    \langle \Delta E_{\rm exc} \rangle = n_{\rm exc}^1 \cdot E_1 + n_{\rm exc}^2 \cdot E_2 \text{, }
\end{equation}
\begin{equation}
    \label{eq:app_gaussian_excitation_variance}
    \sigma_{\Delta E_{\rm exc}}^2 = n_{\rm exc}^1 \cdot E_1^2 + n_{\rm exc}^2 \cdot E_2^2 \text{.}
\end{equation}
where \( E_1 \) and \( E_2 \) are the energy levels of the two excitation types and each term represents their mean contribution.
This Gaussian sampling is truncated by resampling negative values or those exceeding \( 2\Delta E_{\text{exc}} \).

When the mean number of excitation collisions is below $n_{\max}^{\rm exc}$, Poisson-distributed values \( p_1 = \text{Poisson}(n_{\rm exc}^1) \)
and \( p_2 = \text{Poisson}(n_{\rm exc}^2) \) are generated.
In case both are positive, the following contribution to the energy-loss is calculated:
\begin{equation}
    \Delta E_{\rm exc} = \left((p_1 + 1) - 2\text{Uniform}(0, 1)\right) \cdot E_1 + \left((p_2 + 1) - 2\text{Uniform}(0, 1)\right) \cdot E_2 \text{,}
\end{equation}
where \( \text{Uniform}(0, 1) \) represents a uniform random variable in the range \([0, 1]\).
This part is added on top of the Poisson distribution to yield a continuous energy-loss distribution
rather than a discrete one.
The unique shape obtained by MC sampling of this model is presented in Panel 2 and 3 of Fig.\ref{fig:pdf_phase_space}.

\subsubsection{Ionization}
\label{sec:appendix_a_ionization}

In the `thick' absorber limit, energy-loss is sampled from a Gaussian distribution with Bethe-Bloch as the mean and Bohr's variance realized by equations \ref{eq:app_bethe_bloch} and~\ref{eq:app_bohr_variance} respectively.
Otherwise, energy-loss is sampled according to the `thin' absorber limit as described in the following section.

We begin by realizing the energy-loss due to a single ionization collision.
It is achieved by applying the inverse sampling method from the density function of Eq.\ref{eq:app_density_function}:
\begin{equation}
u = F(E) = \int_{E_0}^{E} g(x) \, dx\text{, }
\end{equation}
with $ u $ being a uniformly distributed random number within $ [0, 1] $, which leads to:
\begin{equation}
\label{eq:app_e_ion_collision}
\Delta \mathbb{E}_{\rm ion}^{\rm collision} = \frac{E_0}{1 - u\frac{T_{\rm max} - E_0}{T_{\rm max}}}\text{,}
\end{equation}
\begin{equation}
\label{eq:app_exact_total_ionization_loss}
\Delta \mathbb{E}_{\rm ion}^{\rm total} = \sum_{j=1}^{\rm Poisson\left(\langle n_3 \rangle \right)} \Delta \mathbb{E}_{\rm ion}^{\rm collision}\text{,}
\end{equation}
where $\Delta \mathbb{E}_{\rm ion}^{\rm total}$ is the total energy-loss due to ionization collisions.
This kind of sampling is identified as a `compound Poisson process', which is extensively described in App.~\ref{sec:appendix_b} and~\ref{sec:appendix_c}.
To extract only the continuous energy-loss $T_{\rm max}$ should be replaced with $T_{\rm up}$.

For $\langle n_3 \rangle$ larger than $n_{\max}^{\rm cont}$ (fixed to be equal to 8), the modified-Urb\'an model approximates the total energy-loss to reduce computation time according to the rational explained in section~\ref{subsec:thin_absorber}.

The energy-loss due to hard collisions, with energy transfer between $w_3$ and $T_{\rm cut}$, are sampled similarly to Eq.~\ref{eq:app_exact_total_ionization_loss} with $E_0$ in the numerator, $T_{\rm max}$, $\langle n_3 \rangle$, and $E_0$ in the denominator updated to $w_3$, $T_{\rm cut}$ and $\langle n_3 \rangle - \langle n_a \rangle$, respectively.
$\langle n_a \rangle$ stands for the mean number of soft ionization collisions and $w_3$ is the dynamic energy threshold distinguishing soft and hard collisions and they are implemented as following:
\begin{equation}
    w_3=\alpha E_0 = \frac{\frac{T_{\rm cut}}{E_0}(n_{\max}^{\rm cont} + \langle n_3 \rangle)}{\frac{T_{\rm cut}}{E_0}n_{\max}^{\rm cont} + \langle n_3 \rangle}E_0 \text{,}
\end{equation}
\begin{equation}
    \langle n_a \rangle = \langle n_3 \rangle \frac{T_{\rm cut}}{E_0}\frac{\alpha-1}{\left( \frac{T_{\rm cut}}{E_0} - 1 \right)\alpha} \text{.}
\end{equation}

The energy-loss due to soft collisions, with energy transfer below $w_3$ (yet above $E_0$) are sampled from a Gaussian with mean and variance calculated as:
\begin{equation}
\langle \Delta E_{\rm ion}^{\rm soft} \rangle = \langle n_a \rangle \alpha_1E_0, \quad \sigma_{\rm soft}^2 = \langle n_a \rangle \left( \alpha -\alpha_1^2 \right)E_0^2 \text{, }
\end{equation}
where $\alpha_1=\frac{\alpha \ln(\alpha)}{\alpha - 1}$.

Finally, the energy-loss due to continuous ionization is realized as following:
\begin{equation}
\label{eq:app_continuous_ionization}
\Delta E_{\text{ion}}^{\text{cnt}} =
\begin{cases}
\sum_{j=1}^{\rm Pois.\left(\langle n_3 \rangle \right)} \frac{E_0}{1 - u_j\frac{T_{\rm cut} - E_0}{T_{\rm cut}}},
& \text{ if } \langle n_3 \rangle < n_{\max}^{\rm cont} \\
\sum_{j=1}^{\rm Pois.\left(\langle n_3 \rangle - \langle n_a \rangle \right)} \frac{w_3}{1 - u_j\frac{T_{\rm cut} - w_3}{T_{\rm cut}}} + \text{Gaus.} \left( \langle \Delta E_{\rm ion}^{\rm soft} \rangle, \sigma_{\rm soft} \right),
& \text{ otherwise.}\\
\end{cases}
\end{equation}

The sampling of energy-loss for secondary electron production employs a rejection sampling strategy. The rationale for this approach is that the full differential cross-section for secondary production, which includes relativistic spin-dependent corrections, is mathematically complex and difficult to invert analytically.
To overcome this, the algorithm utilizes a two-step process.

First, a candidate energy transfer is sampled by taking the integration rang of Eq.~\ref{eq:app_density_function} as $[T_{\text{cut}}, T_{\max}]$, leading to:
\begin{equation}
\Delta E_{\rm ion}^{\rm sec} = \frac{T_{\max}}{1 - u\left(1 - \frac{T_{\max}}{T_{\text{cut}}}\right)}\text{. }\label{eq:app_un_trunc_secondary_loss}
\end{equation}

Second, to account for the specific relativistic kinematics that deviate from the pure Rutherford scattering, this candidate is passed through a rejection filter. The acceptance probability is defined as:
\begin{equation}
\label{eq:app_p_accept}
    P_{\rm accept} = 1 - \beta^2 \left( \frac{\Delta E_{\rm ion}^{\rm sec}}{T_{\max}} \right)
\end{equation}
In the Monte Carlo loop, a uniform random variable $q \in [0,1]$ is generated; the candidate energy is accepted if $q < P_{\rm accept}$ and rejected otherwise. This efficiently transforms the approximate Rutherford distribution into the precise physical distribution required by the model without requiring expensive numerical integration.

\subsubsection{Summary of Monte Carlo processes}
The total energy-loss of a single step is given by:
\begin{equation}
\Delta E = \Delta E_{\text{exc}} + \Delta E_{\text{ion}}^{\text{cnt}} + \Delta E_{\text{ion}}^{\text{sec}}\text{, }
\end{equation}
with the fluctuation in energy-loss resulting from variations in excitation and ionization processes.

It is noteable that the processes are partitioned to regimes as following:
\begin{itemize}
    \item
        When the mean energy-loss, $\langle \frac{dE}{dx} \rangle \cdot L$ is below $10$~eV, the energy-loss is simply
        the mean loss.
    \item
        In the thick limit, the energy-loss is sampled from a Gaussian distribution with Bohr's variance or the Gamma
        distribution.
    \item
        In the thin limit, continuous excitation and ionization processes are sampled separately and combined to yield the total energy-loss.
        The excitation part is sampled from a combination of Poisson and uniform distributions for low excitation collision rate, or gaussian distribution for high.
        The ionization part is sampled from a `compound Poisson process', with an additional gaussian fluctuation for large mean number of collisions.
    \item
        A step is terminated when energy transfer at a specific collision, from primary particle to atomic electron, exceeds $T_{\rm cut}$.
        This yields energy-loss of the primary particle, which is sampled through a rejection filter algorithm of a transformed uniform variable.
        A particle that reaches an energy below $T_{\rm cut}$, cannot generate a secondary particle.
        In this case, the step would terminate based on the trajectory's termination criteria.
\end{itemize}

\clearpage

\section{Mathematical derivation of the straggling functions}\label{sec:appendix_b}
\setlength{\parskip}{1em}
\setlength{\parindent}{0.5em}

We derive here, for the first time, an analytical expression for the straggling function of GEANT4's fluctuation model. 
Our derivation targets the sampling algorithms requiring non-trivial PDF formulations: excitation and continuous ionization at low collision rates, as well as the rejection loop for secondary generation.
The total energy-loss distribution is rigorously accounted for by convolving these (and the trivial) process-specific PDFs.

While GEANT4 standardly employs Gaussian approximations (via the Central Limit Theorem) at high collision rates to optimize computational efficiency, our derivations allow us to bypass this approximation. By extending the rigorous low-collision formulas to high-rate regimes, we can calculate the 'exact' straggling function of the model. Furthermore, our framework remains flexible enough to reproduce the standard approximated PDFs used by GEANT4 when desired.

\subsection{Excitation}\label{subsec:continuous_excitations_pdf}

energy-loss through excitation at high collision rate is simply approximated by a Gaussian.
At low rate, excitation in the \texttt{G4UniversalFluctuation} model is represented as:
\begin{equation}
\label{eq:app_e_exc_sampling}
\Delta E_{\text{exc}} \propto \text{Poisson}(\lambda) + 1 - 2 \cdot \text{Uniform}(0,1)\text{.}
\end{equation}

To derive the PDF of this energy-loss, we start by analyzing the components of the expression
and then combining them by a simple convolution.

The PDF of a Poisson-distributed random variable shifted by $+1$ can be expressed as:
\begin{equation}
g(\xi) = P(\xi + 1 = k) = P(\xi = k - 1) = \frac{e^{-\lambda} \lambda^{k-1}}{(k-1)!}\text{.}
\end{equation}

The second component, $-2 \cdot \text{Uniform}(0, 1)$, has a uniform PDF over the interval $[-2, 0)$, described by:
\begin{equation}
f(x) =
\begin{cases}
0 & \text{if } x < -2 \\
\frac{1}{2} & \text{if } -2 \leq x < 0 \\
0 & \text{if } x \geq 0
\end{cases}
\end{equation}

The total energy-loss due to excitation can be expressed as the sum of these two independent random variables.
The PDF of this sum is obtained by a convolution of $f$ with $g$ to $(f * g)(x)$, which we denote as $\mathbb{P}_{\Delta E_{\rm exc}}(x)$.

For our specific PDFs, the convolution can be calculated as follows:
\begin{equation}
\mathbb{P}_{\Delta E_{\rm exc}}(x) = \sum_{k} \frac{e^{-\lambda} \lambda^{k-1}}{(k-1)!} f(x - k)\text{, }
\end{equation}
where $\lambda$ ($k$) is the mean (actual) number of excitations and:
\begin{equation}
f(x - k) =
\begin{cases}
0 & \text{if } x < k - 2 \\
\frac{1}{2} & \text{if } k - 2 \leq x < k \\
0 & \text{if } x \geq k \text{.}
\end{cases}
\end{equation}

Thus, this convolution reduces to two primary terms, evaluated when $x = k-1$ or $x = k$, with $k$ as an integer.
Such that:
\begin{equation}
\label{eq:app_e_exc_pdf}
\mathbb{P}_{\Delta E_{\rm exc}}(x) = \frac{e^{-\lambda}}{2}\left(\frac{\lambda^{\lfloor x \rfloor}}{\lfloor x \rfloor!} + \frac{\lambda^{\lfloor x \rfloor + 1}}{(\lfloor x \rfloor + 1)!}\right),
\end{equation}
where $x \geq 0$ and $\lfloor x \rfloor$ denotes the floor function of $x$.
See panels 2 and 3 in Fig.~\ref{fig:pdf_phase_space}.

\subsection{Continuous ionization}
\label{subsec:continuous_ionization_pdf}
The formulation of Eq.~\ref{eq:app_exact_total_ionization_loss} is identified as a `compound Poisson process', which is
a stochastic process representing the sum of a Poisson number of independent and identically distributed random variables.
The compound Poisson process and its output `characteristic function' are outlined in Sec.~\ref{sec:appendix_c}.
We use the characteristic function methodology to derive its PDF\@.
The full process is detailed below.

A characteristic function of a random variable $X$ is defined as the expected value of $e^{\mathbbm{i}tX}$:
\begin{equation}
\label{eq:app_characteristic_function}
\eta_X(t) = \mathbb{E}\left[ e^{\mathbbm{i}tX} \right] = \int_{-\infty}^{\infty} e^{\mathbbm{i}tx} f(x) \, dx \text{, }
\end{equation}
where $f(x)$ is the PDF of $X$.

\ref{sec:appendix_c} presents a rigorous derivation of the `characteristic function' for the compound Poisson process, leading to the PDF of the full energy-loss due to ionization:
\begin{equation}
\label{eq:app_e_ion_characteristic}
\varphi_{\Delta \mathbb{E}_{\rm ion}^{\rm total}}(t) = \exp \left( \lambda \left( \phi_{\Delta \mathbb{E}_{\rm ion}^{\rm coll.}}(t) - 1 \right) \right)\text{, }
\end{equation}
where $\lambda$ is the mean number of ionizations and $\phi_{\Delta \mathbb{E}_{\rm ion}^{\rm coll.}}(t)$ is the characteristic function of $\Delta \mathbb{E}_{\rm ion}^{\rm coll.}$ defined in Eq.\ref{eq:app_e_ion_collision}.
To get this characteristic function, we first need to derive $f(x)$, which can be obtained from the CDF of $X$:
\begin{equation}
  F(x) = P(\Delta \mathbb{E}_{\rm ion}^{\rm coll.} < x) = P\left( \frac{w_3}{1 - w\xi} < x \right) = P\left( \xi < \frac{x - w_3}{x w} \right) = \frac{x - w_3}{x w}\text{,}
\end{equation}
where $w$ and $w_3$ are realized by the normalization parameters of Eq.~\ref{eq:app_density_function} with $w \equiv (T_{\rm cut} - w_3)/T_{\rm cut}$.
Since \(\xi\) is uniform, the probability \(P(\xi < u)\) is simply $u$ for \(u \in (0,1)\), which yields:
\begin{equation}
F(x) = \frac{x - w_3}{x w}\text{.}
\end{equation}

We get $f(x)$ by differentiating \(F(x)\) with respect to \(x\):
\begin{equation}
\label{eq:app_e_i_ion_pdf}
f(x) = F'(x) = \frac{w_3}{w x^2}\text{.}
\end{equation}

The characteristic function of $\Delta \mathbb{E}_{\rm ion}^{\rm coll.}$ is given from the definition in Eq.~\ref{eq:app_characteristic_function} as:
\begin{equation}
\phi_{\Delta \mathbb{E}_{\rm ion}^{\rm coll.}}(t) = \mathbb{E}\left[ e^{\mathbbm{i}t \Delta \mathbb{E}_{\rm ion}^{\rm coll.}} \right] = \int_{w_3}^{T_{\rm cut}} e^{\mathbbm{i}tx} f(x) \, dx
= \frac{w_3}{w} \int_{w_3}^{T_{\rm cut}} \frac{e^{\mathbbm{i}tx}}{x^2} \, dx\text{, }
\end{equation}
where \(\left[ w_3, T_{\rm cut} \right]\) is the range of $\Delta \mathbb{E}_{\rm ion}^{\rm coll.}$, which is written as $x$ in the integral for convenience.

Separating the integral into its real and imaginary components gives:
\begin{equation}
\label{eq:app_e_i_ion_characteristic}
\phi_{\Delta \mathbb{E}_{\rm ion}^{\rm coll.}}(t) = \frac{w_3}{w} \left( \int_{w_3}^{T_{\rm cut}} \frac{\cos(tx)}{x^2} \, dx + \mathbbm{i} \int_{w_3}^{T_{\rm cut}} \frac{\sin(tx)}{x^2} \, dx \right)\text{.}
\end{equation}

Utilizing integration by parts, we can express the integrals in terms of
sine and cosine integrals (\texttt{Si} and \texttt{Ci} functions):
\begin{equation}
\label{eq:app_cosine_integral_parts}
\int\limits_{w_3}^{T_{\rm cut}} \frac{\cos{(ty)}}{y^2} \, dy
= \frac{\cos(w_3 t)}{w_3} - \frac{\cos\left(T_{\rm cut} t\right)}{T_{\rm cut}} + t \, { Si}(w_3 t) - t \, { Si}\left(T_{\rm cut} t\right)
\end{equation}
and
\begin{equation}
\label{eq:app_sine_integral_parts}
\int\limits_{w_3}^{T_{\rm cut}} \frac{\sin{(ty)}}{y^2} \, dy
= T_{\rm cut} t - \frac{\sin\left(T_{\rm cut} t\right)}{T_{\rm cut}} - t \, { Ci}(w_3 t) + t \, { Ci}\left(T_{\rm cut} t\right)
\end{equation}

\begin{figure}[!ht]
    \centering
    \includegraphics[width=\textwidth]{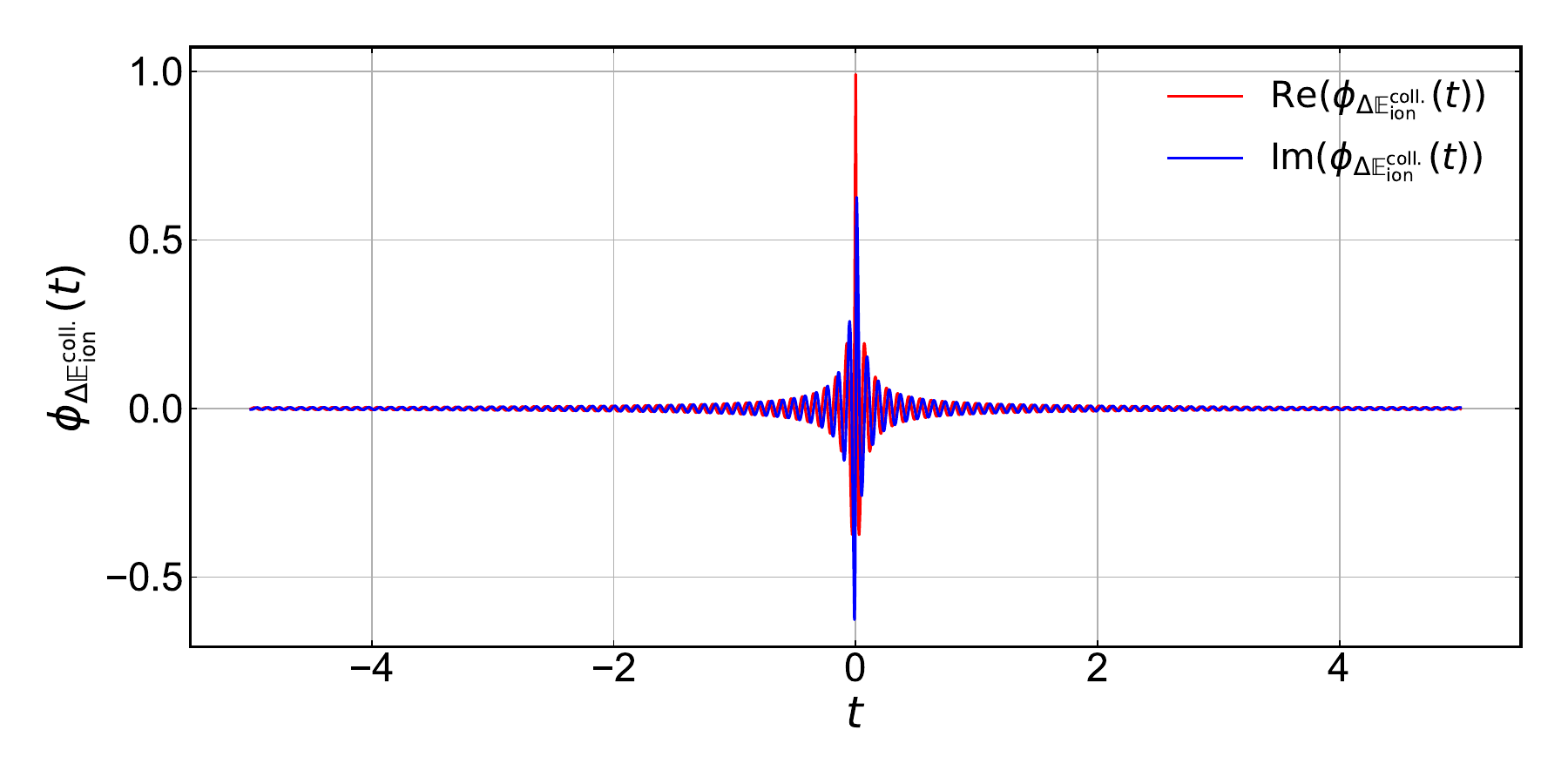}
    \caption{
    The characteristic function of $\Delta \mathbb{E}_{\rm ion}^{\rm coll.}$ for a $10~\text{MeV}$ proton travelling $1~\mu \text{m}$ through
        aluminum as derived from Eq.~\ref{eq:app_e_i_ion_full_characteristic_expression}.
        The red (blue) curve is the real (imaginary) part of the function.
    }
    \label{fig:e_i_ion_characteristic}
    \hfill
\end{figure}

Finally, the characteristic function for $\Delta \mathbb{E}_{\rm ion}^{\rm coll.}$ can be written as:
\begin{align}
\label{eq:app_e_i_ion_full_characteristic_expression}
\phi_{\Delta \mathbb{E}_{\rm ion}^{\rm coll.}}(t) = \frac{\cos(w_3 t)}{w_3} - \frac{\cos\left(T_{\rm cut} t\right)}{T_{\rm cut}} + t \,
{ Si}(w_3 t) - t \, { Si}\left(T_{\rm cut} t\right) \notag \\
+ \mathbbm{i} \left( \frac{\sin(w_3 t)}{w_3} -
\frac{\sin\left(T_{\rm cut} t\right)}{T_{\rm cut}} - t \,
{ Ci}(w_3 t) + t \, { Ci}\left(T_{\rm cut} t\right) \right)\text{.}
\end{align}
The real and imaginary components of $\phi_{\Delta \mathbb{E}_{\rm ion}^{\rm coll.}}(t)$ are plotted in Fig.~\ref{fig:e_i_ion_characteristic}.

These integrals can be evaluated numerically or by using built-in special functions for sine
and cosine integrals.

\begin{figure}[!ht]
    \centering
    \includegraphics[width=\textwidth]{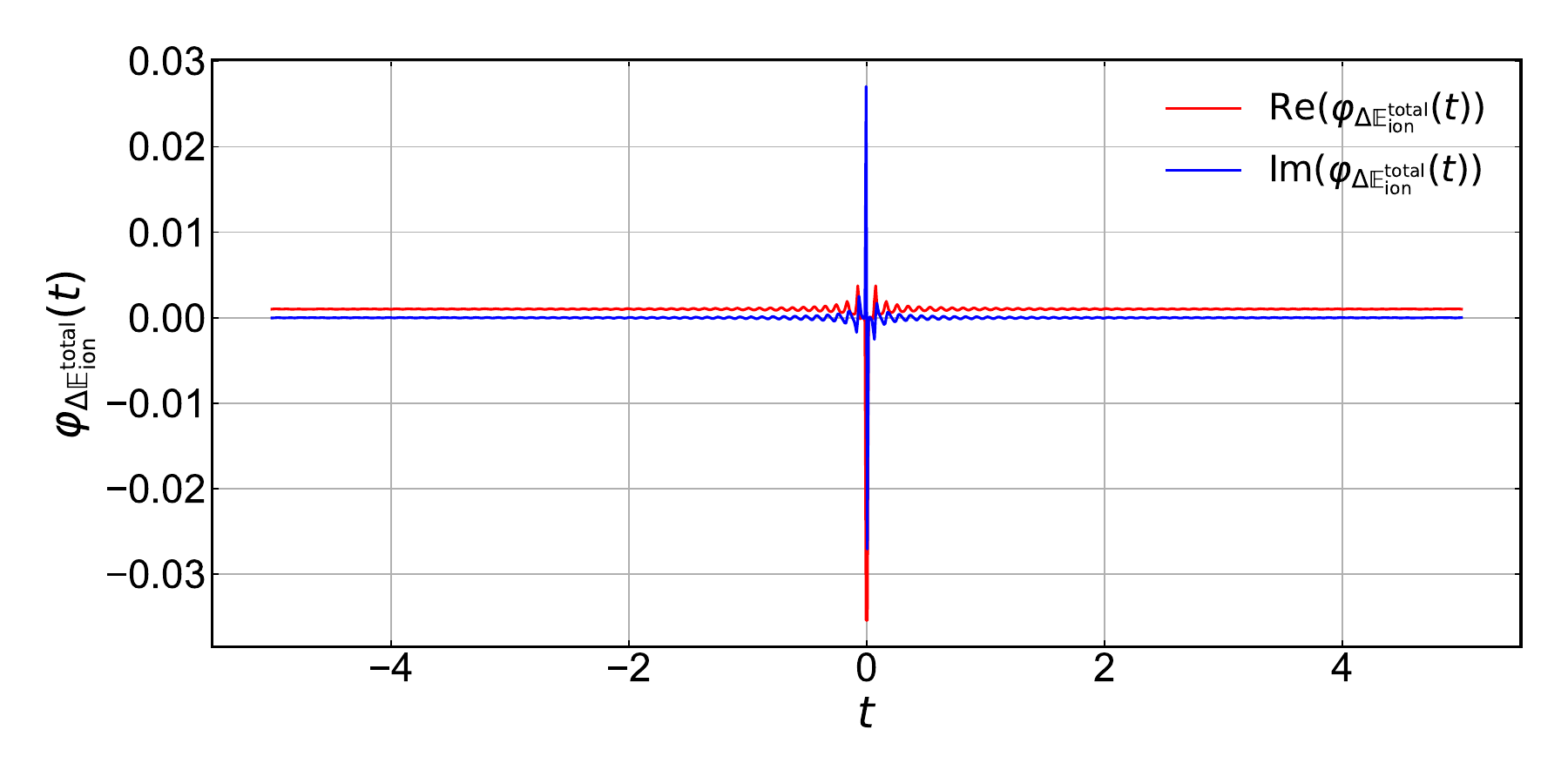}
    \caption{
    The characteristic function of $\Delta \mathbb{E}_{\rm ion}^{\rm total}$ for a $10~\text{MeV}$ proton travelling $1~\mu \text{m}$ through
        aluminum as derived from equations~\ref{eq:app_e_ion_characteristic}
        and~\ref{eq:app_e_i_ion_full_characteristic_expression}. The red (blue) curve is the real (imaginary) part of the function.
    }
    \label{fig:e_ion_characteristic}
    \hfill
\end{figure}

The real and imaginary components of \(\varphi_{\Delta \mathbb{E}_{\rm ion}^{\rm total}}(t)\), which is computed by substituting
Eq.~\ref{eq:app_e_i_ion_full_characteristic_expression} into Eq.~\ref{eq:app_e_ion_characteristic},
are plotted in Fig.~\ref{fig:e_ion_characteristic}.
It can be seen that the bare shapes are highly non-trivial and hence so is their integral.

By identifying the characteristic function (defined in Eq.~\ref{eq:app_characteristic_function}) as a Fourier transformation,
the PDF can be simply computed by the inverse Fourier transform of \(\varphi_{\Delta \mathbb{E}_{\rm ion}^{\rm total}}(t)\):
\begin{equation}
\mathbb{P}_{\Delta \mathbb{E}_{\rm ion}^{\rm total}}(x) = \frac{1}{2\pi} \int_{-\infty}^{\infty} e^{-\mathbbm{i}tx} \varphi_{\Delta \mathbb{E}_{\rm ion}^{\rm total}}(t) \, dt
=\frac{1}{2\pi} \int_{-\infty}^{\infty} e^{-\mathbbm{i}tx} \exp \left( \lambda \left( \phi_{\Delta \mathbb{E}_{\rm ion}^{\rm coll.}}(t) - 1 \right) \right) \, dt\text{.}
\end{equation}

Substituting $\phi_{\Delta \mathbb{E}_{\rm ion}^{\rm coll.}}(t)$ with the expression presented in Eq.~\ref{eq:app_e_i_ion_full_characteristic_expression}
yields the full formulation of $\mathbb{P}_{\Delta \mathbb{E}_{\rm ion}^{\rm total}}(x)$, which is unsolvable analytically.

In practice, the inverse Fourier transform can be approximated using
numerical methods such as the discrete Fourier transform (DFT). The DFT
approximates the integral as a sum over discrete points:
\begin{equation}
\label{eq:app_e_ion_pdf}
\mathbb{P}_{\Delta \mathbb{E}_{\rm ion}^{\rm total}}(x) = \frac{1}{2\pi} \sum_{k=0}^{N-1} e^{-\mathbbm{i}xk\Delta t} \exp \left( \lambda \left( \phi_{\Delta \mathbb{E}_{\rm ion}^{\rm coll.}}(k\Delta t) - 1 \right) \right) \Delta t\text{, }
\end{equation}
where $N$ is the number of discrete points, and $\Delta t$ is the
spacing between points.
This numerical approach allows for the rapid computation of the PDF of $\Delta \mathbb{E}_{\rm ion}^{\rm total}$
and if $N$ is large enough it returns an extremely accurate representation of the actual integral result.
See Fig.~\ref{fig:pdf_phase_space}.

The PDF of the compound Poisson process $\Delta \mathbb{E}_{\rm ion}^{\rm total}$ discussed above is written and calculated for
the first time.
It can be seen that the behaviour is extremely non-trivial and cannot be approximated e.g, with a Landau distribution.
Furthermore, we see that this derivation works extremely well across a huge dynamic range of several orders of magnitude
in step lengths and kinetic energies.
Finally, while increasing $N$ indefinitely will result in a growing precision of the final integral estimate, the
calculation time will increasingly grow in parallel.
We have found that $N$ can be determined for any point in the relevant phase-space using a few simple rules derived from
the behaviour of the characteristic function from Eq.~\ref{eq:app_e_ion_characteristic} (see Fig.~\ref{fig:e_ion_characteristic} for example).
With this, we see that the average time necessary to calculate the full expression of Eq.~\ref{eq:app_e_ion_pdf} at the desired precision for
any point in the phase-space is approximately only 1-3~seconds on a regular personal computer.

\subsection{Combination of all continuous loss processes}\label{subsec:combination-of-all-continuous-loss-processes}

The sampling of \(\Delta E_{\text{exc}}\) and \(\Delta E_{\text{ion}}^{\text{cnt}}\) are divided (each) into two regions, as described
in section~\ref{subsubsec:excitations} and Eq.~\ref{eq:app_continuous_ionization}, respectively.
The total energy-loss during a particle's `step' is the sum of these two components, ensuring that different
regimes of the underlying physics are captured.
The respective PDFs derived in this section are convolved to obtain the PDF for the total energy-loss in each regime.

So far, we haven't dealt with the case where the Poisson number of either excitation or ionization process is zero (no collisions).
In this case, there is no energy-loss, and the PDF is a Dirac delta function at $\Delta E = 0$, with a probability of
$\alpha$, where alpha is $e^{-\langle n_3 \rangle}$ for low ionization collision rate, $e^{-\left(\langle n_3 \rangle - \langle n_a \rangle\right)}$ for high ionization collision rate and $e^{-\left< n_{\rm exc} \right>}$ for excitation.
The total PDF of the energy-loss is then a mixture of the continuous PDF and the Dirac delta function:
\begin{equation}
    \mathbb{\tilde P}(\Delta E) = (1-\alpha)\mathbb{P}(\Delta E) + \alpha \delta(\Delta E)\text{, }
\end{equation}
where $\mathbb{P}(\Delta E)$ is Eq.~\ref{eq:app_e_exc_pdf} for excitation and Eq.~\ref{eq:app_e_ion_pdf} for ionization.
This normalization procedure has a dramatic effect on the calculation (by convolution) of the combined PDF of the continuous
loss processes, and it should be performed for the PDF of each individual process before the convolution.

\subsection{Ionization with physical production of secondary electrons} \label{subsec:secondary_ionization}
Finally, we derive the straggling function for the ionization component responsible for secondary electron production. 
In the GEANT4 architecture, this process is simulated via a rejection sampling method to accurately reproduce the full differential cross-section.
Translating this algorithmic procedure into a PDF requires a rigorous statistical treatment of the acceptance-rejection loop.
Below, we derive the PDFs for both the candidate energy and the acceptance variable, convolving them to produce the final analytical expression for energy-loss in this regime.

The following identity is employed for a random variable $\xi$ with PDF $f_{\xi}$ and a transformed variable $\eta = g(\xi)$ with PDF $f_{\eta}$:
\begin{equation}
f_{\eta}(x) = f_{\xi}(g^{-1}(x)) \left| \frac{d}{dx} g(x) \right|\text{, } \label{eq:app_pdf_transformation}
\end{equation}
where $g^{-1}$ is the inverse function of $g$.
For simplicity, we will refer to $P_{\rm accept}$ defined in Eq.~\ref{eq:app_p_accept}, which is a random variable, as y hereafter.

Isolating $\Delta E_{\rm ion}^{\rm sec}$ in Eq.~\ref{eq:app_p_accept} and differentiating it with respect to $y$ gives:
\begin{equation}
g^{-1}(x) = \Delta E_{\text{ion}}^{\text{sec}} = \frac{T_{\max}}{\beta^2}\left(1 - x\right),\label{eq:app_truncated_secondary_loss_vs_f_n}
\end{equation}
and
\begin{equation}
\left| \frac{d}{dx} g(x) \right| = \left| \frac{d \Delta E_{\text{ion}}^{\text{sec}}}{dx} \right| = \frac{T_{\max}}{\beta^2}.\label{eq:app_truncated_secondary_loss_derivative}
\end{equation}

Substituting equations~\ref{eq:app_truncated_secondary_loss_vs_f_n} and~\ref{eq:app_truncated_secondary_loss_derivative} into Eq.~\ref{eq:app_pdf_transformation} results in:
\begin{equation}
    f_y(x) = \frac{T_{\max}}{\left( \frac{T_{\max}}{T_{\rm cut}} - 1 \right) (\frac{T_{\max}}{\beta^2}\left(1 - x\right))^2} \frac{T_{\max}}{\beta^2} \\
    = \frac{\beta^2}{\left( \frac{T_{\max}}{T_{\rm cut}} - 1 \right)\left(1 - x\right)^2}.\label{eq:app_loop_variable_pdf}
\end{equation}

In the MC procedure, the loop at which $u$ is sampled and $y$ is computed, as defined in Eq.~\ref{eq:app_p_accept},
is repeated until $y_n > u_n$, where $u$ is a uniformly distributed random variable in $[0, 1]$.
The probability of satisfying this condition is given by:
\begin{equation}
\label{eq:app_loop_ending_probability}
P(u < y) = \int dx \int f_{\text{uy}}(x, y) \, dy = \int f_y(x) \, dx \int_{-\infty}^{x}f_u(y) \, dy = \int P(u < x) f_y(x) \, dx = \int F_u(x) f_y(x) \, dx\text{, }
\end{equation}
where the combined PDF of $u$ and $y$, $f_{\text{uy}}$, is the product of their individual PDFs due to their independence, and $F_u(x) = x$ is the CDF of $u$.

Substituting Eq.~\ref{eq:app_loop_variable_pdf} into Eq.~\ref{eq:app_loop_ending_probability} yields:
\begin{align}
P(u < y) = \frac{\beta^2 T_{\rm cut}}{T_{\max} - T_{\rm cut}} \int_{1 - \beta^2}^{1 - \beta^2 \frac{T_{\rm cut}}{T_{\max}}} \frac{x}{\left(1 - x\right)^2} \, dx
= \frac{\beta^2 T_{\rm cut}}{T_{\max} - T_{\rm cut}} \left[ \ln\left(1 - x\right) - \frac{1}{1 - x}\right]_{1 - \beta^2}^{1 - \beta^2 \frac{T_{\rm cut}}{T_{\max}}} \notag \\
= 1 - \frac{\beta^2 T_{\rm cut}}{T_{\max} - T_{\rm cut}} \ln\left(\frac{T_{\max}}{T_{\rm cut}}\right).
\end{align}

The number of iterations $k$ until the MC loop ends, follows the Bernoulli distribution.
The probability of ending at the $n$-th iteration is, therefore, given by:
\begin{equation}
P(k = n) = P(u < y)(1 - P(u < y))^{(n-1)}.\label{eq:app_bernoulli_distribution}
\end{equation}

For the PDF of $y_n$, which is the result of $n$ trials where $q_i > y_i$ for $i \in [1, n)$, the following expression is considered:
\begin{align}
F(x) \equiv P(y_n < x | u_n < y_n) = \frac{P(y_n < x \bigcap u_n < y_n)}{P(u < y)} \notag \\
= \frac{\int_{1 - \beta^2}^x \int_0^q f_y(q) f_{u}(v) \, dv \, dq}{P(u < y)} = \frac{\int_{1 - \beta^2}^x f_y(q) \left[ \int_0^q f_u(v) \, dv \right] \, dq}{P(u < y)}.\label{eq:app_cdf_f_n}
\end{align}
The term in square brackets is identified as the CDF of $u$, $F_u(q)$, resulting in the PDF of $y_n$:
\begin{equation}
f_{y_n}(x) = \frac{d}{dx} F(x) = \frac{f_y(x) F_u(x)}{P(u < y)}.\label{eq:app_pdf_f_n}
\end{equation}

Using equations~\ref{eq:app_loop_variable_pdf},~\ref{eq:app_loop_ending_probability}, and $F_u(x)=x$ for $0<x<1$, the PDF of $y_n$ becomes:
\begin{equation}
f_{y_n}(x) = \frac{\beta^2}{\left( \frac{T_{\max}}{T_{\rm cut}} - 1 \right) - \beta^2 \ln\left( \frac{T_{\max}}{T_{\rm cut}} \right)} \frac{x}{(1-x)^2}. \label{eq:app_pdf_f_n_final}
\end{equation}

\begin{figure}[!ht]
    \centering
    \includegraphics[width=1\textwidth]{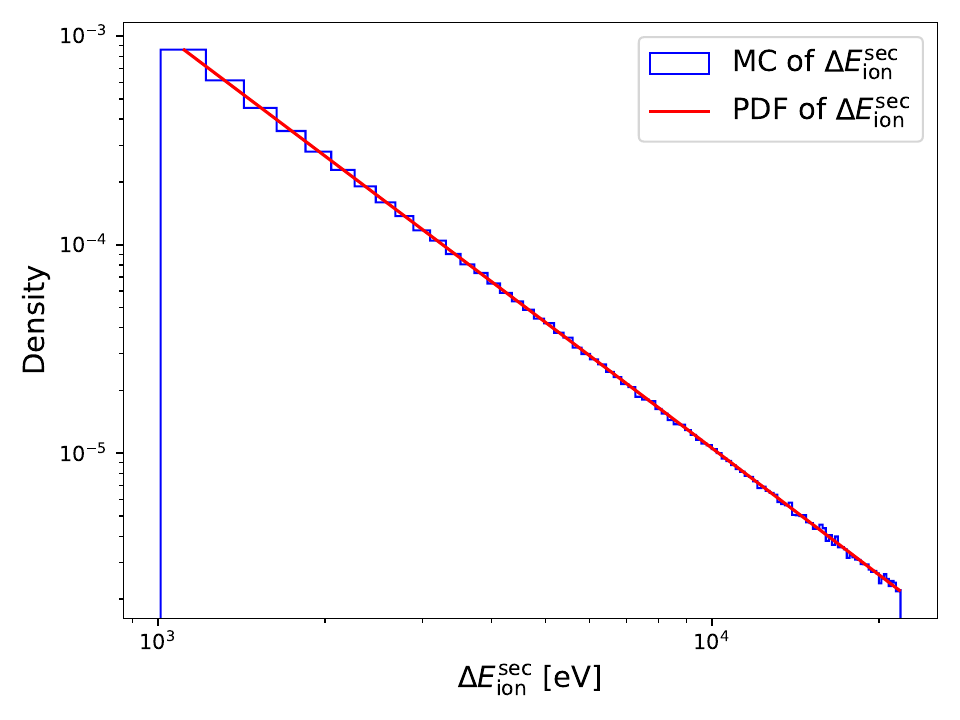}
    \caption{
    The PDF of the energy-loss via physical secondary electrons production for a $10~\text{MeV}$ proton traveling through aluminum (we note again that there is no dependence of the step length).
    The red curve is the derived PDF from Eq.~\ref{eq:app_discrete_ionization_pdf}, while the blue histogram is the
    MC simulation of 1M steps from Eq.~\ref{eq:app_un_trunc_secondary_loss} with the rejection criterion from Eq.~\ref{eq:app_p_accept}.
    }
    \label{fig:secondary_energy_loss}
    \hfill
\end{figure}

Finally, using equations~\ref{eq:app_p_accept} and~\ref{eq:app_pdf_transformation}, the PDF of $\Delta E_{\text{ion}}^{\text{sec}}$ is derived:
\begin{equation}
\label{eq:app_discrete_ionization_pdf}
\mathbb{P}_{\Delta E_{\text{ion}}^{\text{sec}}}(x) = \frac{T_{\max} - \beta^2x}{\left(\frac{T_{\max}}{T_{\rm cut}} - 1 - \beta^2 \ln\left(\frac{T_{\max}}{T_{\rm cut}}\right)\right)x^2}\text{.}
\end{equation}
The shape of this distribution for a specific point in the phase-space is illustrated by Fig.~\ref{fig:secondary_energy_loss}.
As it can be seen, while the continuous terms had implicit dependence on the step length via the mean loss parameter, Eq.~\ref{eq:app_discrete_ionization_pdf} has no such dependency.

\clearpage

\section{Compound Poisson Process}\label{sec:appendix_c}
\setlength{\parskip}{1em}
\setlength{\parindent}{0.5em}

\numberwithin{equation}{section}
\setcounter{equation}{0} 

The compound Poisson process is a stochastic process where the total sum
of random variables is modeled when the count of these variables follows
a Poisson distribution.

Let:
\begin{itemize}
    \item \(N\) be the number of ionization collisions in a given period, following a Poisson distribution with mean \(\lambda\).

    \item $\Delta \mathbb{E}_{\rm ion}^{\rm coll.}$ is an independent and identically distributed random variable of the energy-loss associated with a single collision.
\end{itemize}
Then, the total energy-loss through ionization is defined as:
\begin{equation}
\Delta \mathbb{E}_{\rm ion}^{\rm total} = \sum_{j=1}^{N} \mathbb{E}_{\rm ion}^{\rm coll.}\text{.}
\end{equation}

The characteristic function of $\Delta \mathbb{E}_{\rm ion}^{\rm total}$, denoted
\(\varphi_{\Delta \mathbb{E}_{\rm ion}^{\rm total}}(t)\), is defined as the expectation value of the $e^{\mathbbm{i}t\Delta \mathbb{E}_{\rm ion}^{\rm total}}$:
\begin{equation}
\varphi_{\Delta \mathbb{E}_{\rm ion}^{\rm total}}(t) = \mathbb{E}\left[ e^{\mathbbm{i}t \Delta \mathbb{E}_{\rm ion}^{\rm total}} \right] \text{,}
\end{equation}
where the expectation value can be expressed as an infinite sum.

Given \(N = n\), where n is the number of collisions, the total energy-loss \(\Delta \mathbb{E}_{\rm ion}^{\rm total}\) is the sum of \(n\) independent and identically distributed random variables $\Delta \mathbb{E}_{\rm ion}^{\rm coll.}$.
Thus, we can rewrite the characteristic function as a conditional expectation:
\begin{equation}
\varphi_{\Delta \mathbb{E}_{\rm ion}^{\rm total}}(t) = \sum_{n=0}^{\infty} \mathbb{E} \left[ e^{\mathbbm{i}t \sum_{j=1}^{n} \Delta \mathbb{E}_{\rm ion}^{\rm coll.}} \right] \frac{\lambda^n e^{-\lambda}}{n!} \text{, }
\end{equation}
where $\lambda$ is the Poisson parameter and the mean number of ionization collisions.

Since $\Delta \mathbb{E}_{\rm ion}^{\rm coll.}$ are independent and identically distributed, the characteristic function for
the sum \(\sum_{j=1}^{n} \Delta \mathbb{E}_{\rm ion}^{\rm coll.}\) can be simplified to:
\begin{equation}
\mathbb{E} \left[ e^{\mathbbm{i}t \sum_{j=1}^{n} \Delta \mathbb{E}_{\rm ion}^{\rm coll.}} \right] = \left( \phi_{\Delta \mathbb{E}_{\rm ion}^{\rm coll.}}(t) \right)^n
\end{equation}
which yields:
\begin{equation}
\varphi_{\Delta \mathbb{E}_{\rm ion}^{\rm total}}(t) = \sum_{n=0}^{\infty} \left( \phi_{\Delta \mathbb{E}_{\rm ion}^{\rm coll.}}(t) \right)^n \frac{\lambda^n e^{-\lambda}}{n!}
\end{equation}

The series now resembles the Taylor expansion of the exponential
function \(e^{x} = \sum_{n=0}^{\infty} \frac{x^n}{n!}\).
Applying this identity, we get:
\begin{equation}
\varphi_{\Delta E_{\text{ion}}^{\text{cnt}}}(t) = e^{-\lambda} \sum_{n=0}^{\infty} \frac{\left( \lambda \phi_{\Delta \mathbb{E}_{\rm ion}^{\rm coll.}}(t) \right)^n}{n!} = e^{-\lambda} e^{\lambda \phi_{\Delta \mathbb{E}_{\rm ion}^{\rm coll.}}(t)}
\end{equation}
Simplifying, we find:
\begin{equation}
\varphi_{\Delta \mathbb{E}_{\rm ion}^{\rm total}}(t) = \exp \left( \lambda \left( \phi_{\Delta \mathbb{E}_{\rm ion}^{\rm coll.}}(t) - 1 \right) \right)
\end{equation}
where \(\phi_{\Delta \mathbb{E}_{\rm ion}^{\rm coll.}}(t)\) is the characteristic function of the energy-loss per individual collision.
This function can be used to describe the distribution of $\Delta \mathbb{E}_{\rm ion}^{\rm total}$ analytically.

\clearpage

\section{Neural Networks}\label{sec:appendix_d}
\setlength{\parskip}{1em}
\setlength{\parindent}{0.5em}

\subsection{Feature Scaling}
The features scaling of each module is performed as following:

\noindent\textbf{Continuous: }
\begin{itemize}
    \item $E \to \frac{\rm log(\textit{E}) - min[log(\textit{E})]}{\rm max[log(\textit{E})] - min[log(\textit{E})]}$
    \item $L \to 1- \frac{\rm log(\textit{L}) - min[log(\textit{L})]}{\rm max[log(\textit{L})] - min[log(\textit{L})]}$
\end{itemize}

\noindent\textbf{Secondaries: }
\begin{itemize}
    \item $E \to \frac{\rm log(\textit{E}) - min[log(\textit{E})]}{\rm max[log(\textit{E})] - min[log(\textit{E})]}$
    \item $\theta \to 1 - \frac{\rm log(\theta) - min[log(\theta)]}{\rm max[log(\theta)] - min[log(\theta)]}$
\end{itemize}

\subsection{Architecture}

\begin{figure}[!ht]
    \centering
    \includegraphics[width=0.5\textwidth]{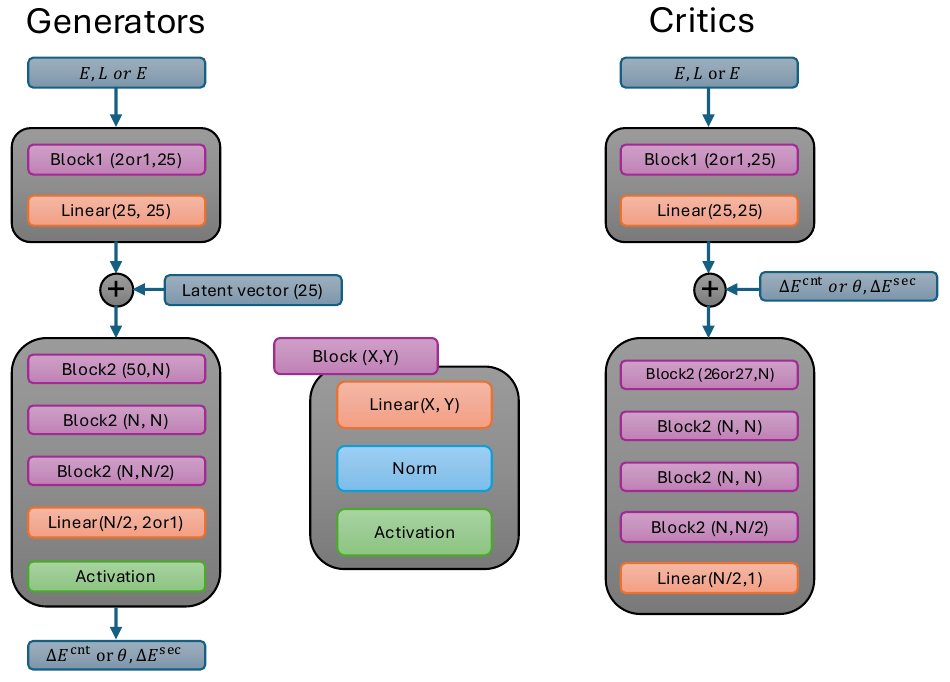}
    \caption{
    The implementation details of the neural networks examined in this paper.
    Block(X, Y) in the purple rectangles is an abbreviation for a repetitive unit of a fully connected linear layer (FCLL) $\to$ normalization $\to$ activation function.
    Below are the normalization and activation layers of the networks.
    PHIN-GAN's generators use LayerNorm and ReLU for both block 1 and 2,
    GAN's generators use InstanceNorm and ReLU for block1 and BatchNorm and ReLU for block2,
    Critics use LayerNorm and ReLU for block1 and LayerNorm and LeakyReLU(0.2) for block2.
    N is 50 (100) for the continuous (secondaries) module.
    The last activation functions of the GAN's continuous and secondaries generators are $1.2\text{Sigmoid}-0.1$ and ReLU respectively.
    The last activation function of the PHIN-GAN's secondaries generator is Hardtanh.
    }
    \label{fig:networks_architecture}
    \hfill
\end{figure}

\section{Figures}\label{sec:appendix_e}
\setlength{\parskip}{1em}
\setlength{\parindent}{0.5em}

\begin{figure}
    \centering
    \includegraphics[width=\textwidth]{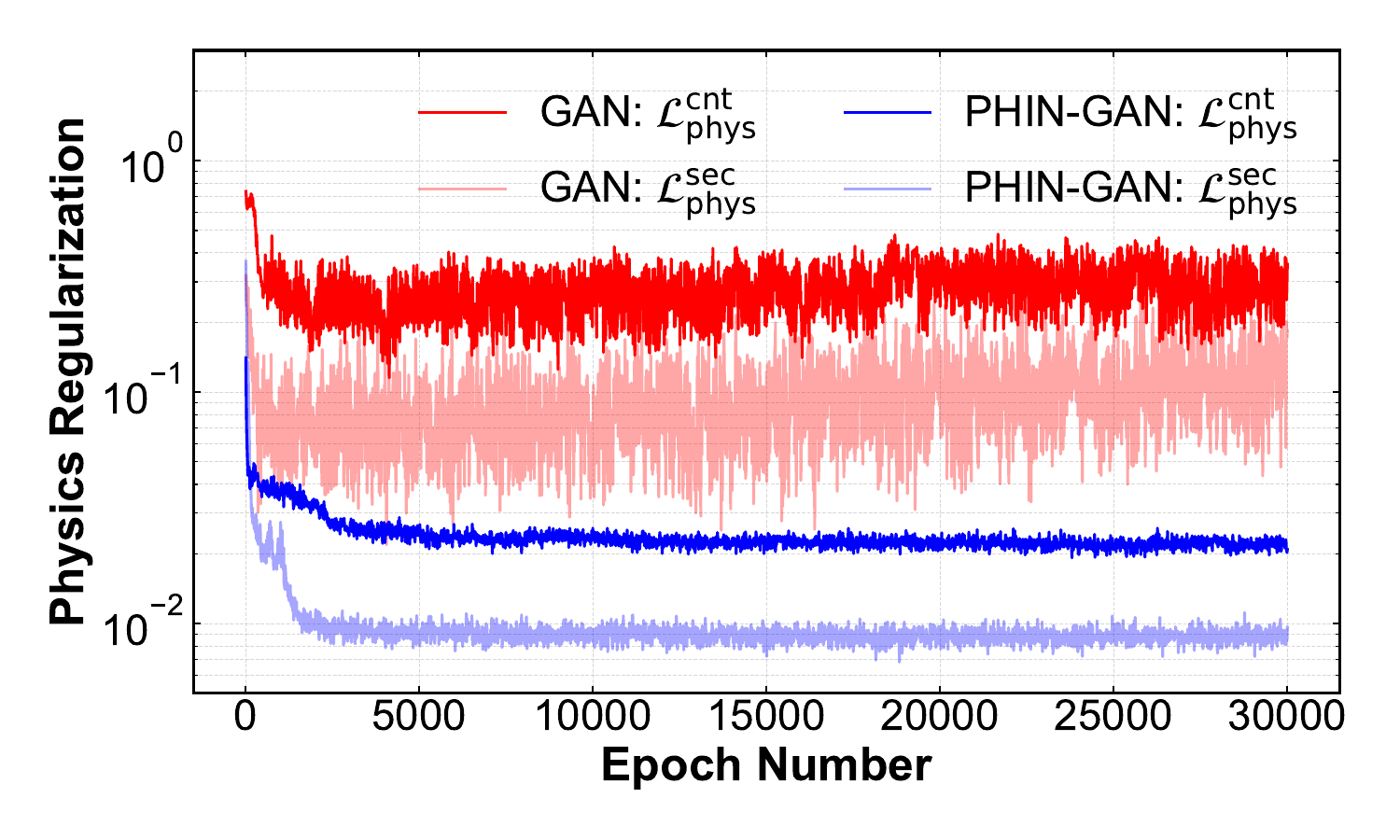}
    \caption{
    The value of the physics regularization along the training session of the GAN and the PHIN-GAN, which is composed of $\mathcal{L}^{\rm sec}_{\rm phys}$ and $\mathcal{L}^{\rm cnt}_{\rm phys}$.
    The values presented here are not multiplied by the weights, such that for the GAN $\alpha^{\rm cnt}=\alpha^{\rm sec}=0$ and for the PHIN-GAN $\alpha^{\rm cnt}=\alpha^{\rm sec}=10^{-1}$.
    }
    \label{fig:losses_vs_epochs}
    \hfill
\end{figure}

\begin{figure}
    \centering
    \includegraphics[width=\textwidth]{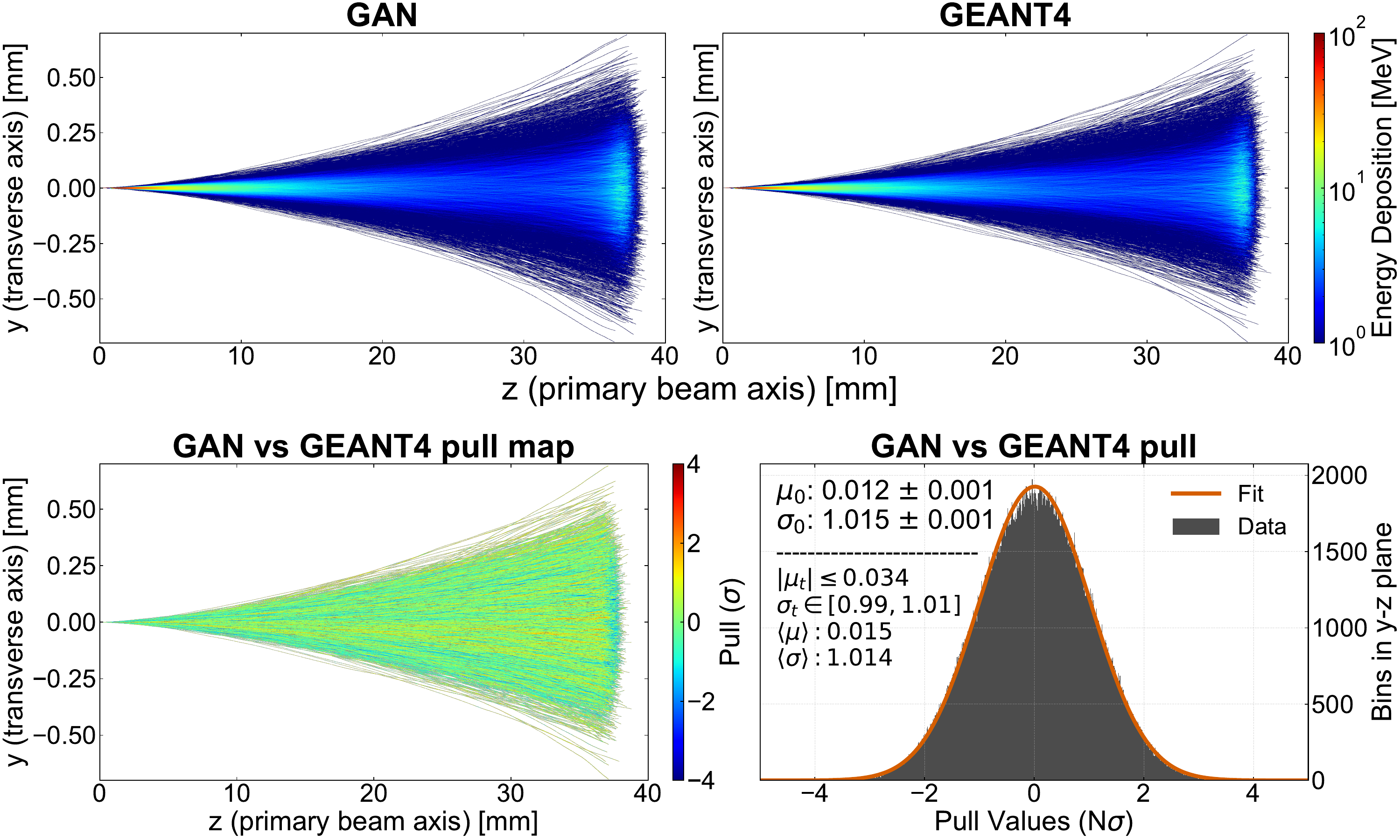}
    \caption{
        A macroscopic comparison of the total energy-loss profiles for two samples, each comprising 10,000 proton trajectories, simulated by GEANT4 and the GAN. 
        The top panels display the binned accumulated total energy-loss of one sample, each.
        The bottom left panel illustrates the spatial map of the calculated pull values between the two samples (from the top panels). 
        The bottom right panel presents the pull value distribution for all $y\textsf{--}z$ bins with more than three entries, overlaid with a Gaussian fit.
        The fit results derived from this specific two-sample comparison are denoted as $\mu_0$ and $\sigma_0$.
        The statistical errors reported by the fit are underestimated (see text) and hence the tolerances for these central values are provided below.
    }
    \label{fig:gan_track_comparison}
    \hfill
\end{figure}


\end{document}